\documentclass{article} 
\usepackage[numbers]{natbib}

\usepackage[final]{neurips_2024}

\usepackage[utf8]{inputenc} 
\usepackage[T1]{fontenc}    
\usepackage{hyperref}       
\usepackage{url}            
\usepackage{booktabs}       
\usepackage{amsfonts}       
\usepackage{nicefrac}       
\usepackage{microtype}      
\usepackage{xcolor}   
\usepackage[table]{xcolor} 
\usepackage{tabularx}       
\usepackage{pifont}
\usepackage{graphicx}
\usepackage{xcolor}
\usepackage{lipsum}
\usepackage{multirow}
\usepackage{marvosym}
\usepackage{xspace}
\usepackage{blindtext}
\usepackage{amsmath}
\usepackage{algorithm}

\usepackage{tcolorbox}
\usepackage{fontawesome5}

\usepackage{algpseudocode}
\usepackage{wrapfig}
\usepackage{enumitem}

\usepackage{subcaption}

\usepackage{amssymb}
\usepackage{amsmath}

\usepackage[table,dvipsnames,svgnames]{xcolor}
\usepackage{tablefootnote}
\usepackage{subcaption}

\usepackage{wrapfig}
\usepackage{lipsum} 
\usepackage{booktabs} 
\usepackage{adjustbox}
\usepackage{graphicx}
\usepackage{placeins}

\definecolor{catgray}{gray}{0.92}
\hypersetup{colorlinks,urlcolor=purple,citecolor=cyan,linkcolor=purple}

\makeatletter
\DeclareRobustCommand\onedot{\futurelet\@let@token\@onedot}
\def\@onedot{\ifx\@let@token.\else.\null\fi\xspace}

\makeatother

\newcommand{\modelname}{\mbox{AuK}\xspace}

\definecolor{citecolor}{rgb}{0.21,0.49,0.74}
\definecolor{linkcolor}{HTML}{ED1C24}
\definecolor{graycolor}{rgb}{0.95,0.95,0.95}
\definecolor{cvprblue}{rgb}{0.21,0.49,0.74}
\definecolor{memory}{rgb}{0.82, 0.51, 0.50} 
\definecolor{current}{rgb}{0.49, 0.60, 0.74} 
\definecolor{softgreen}{HTML}{4CAF50}
\definecolor{softyellow}{HTML}{E0A800}
\definecolor{rowgreen}{HTML}{E8F5E9}
\definecolor{rowyellow}{HTML}{FFF8E1}

\usepackage[capitalize]{cleveref}
\crefname{section}{Sec.}{Secs.}

\crefname{table}{Tab.}{Tabs.}
\crefname{figure}{Fig.}{Figs.}

\usepackage{tikzsymbols}
\definecolor{goodgreen}{RGB}{46, 204, 113}  
\definecolor{normalyellow}{RGB}{241, 196, 15}
\definecolor{badred}{RGB}{231, 76, 60}

\usepackage[toc,page]{appendix}

\title{\modelname Technical Report: An Open-Source Foundational Model \\ for Speech Generation and Editing}

\begin{document}

\maketitle

\makeatletter
\let\original@makefnmark\@makefnmark 
\makeatother

\makeatletter
\def\@makefnmark{}
\makeatother

\makeatletter
\let\@makefnmark\original@makefnmark
\makeatother

\begin{abstract}
We introduce \textbf{\modelname}, an open-source foundational model that unifies speech generation and editing through a common interface of natural-language instructions and audio context. To support this broad capability set, we construct approximately $3.03$ billion instruction--audio instances and $1.95$ million hours of effective supervision across five task families: speech generation, content editing, enhancement and separation, paralinguistic editing, and acoustic editing. \modelname combines a multimodal large language model for semantic conditioning, an VAE jointly trained on speech, general audio, and music for acoustic conditioning, and a hybrid rectified-flow Transformer that performs dual-stream MMDiT blocks followed by unified single-stream DiT blocks for generation. Training begins with generation-only warm-up and proceeds to joint generation--editing pre-training. We then apply complementary post-training strategies: human-feedback preference optimization for open-ended editing and reward-based reinforcement learning for speech generation. To reduce inference cost, we further distill the model with consistency initialization and task-routed Decoupled DMD. The resulting \modelname-Flash performs 4-step inference without classifier-free guidance and achieves a $4.5\times$ wall-clock speedup over the full model under matched conditions. Experiments demonstrate leading performance on zero-shot and instruction-controlled speech generation and general instruction-guided editing, while remaining competitive on signal-level restoration tasks. We release both the source code and model weights to support reproducibility and further research.\end{abstract}

\begin{figure}[h!]
\centering
\includegraphics[width=\textwidth]{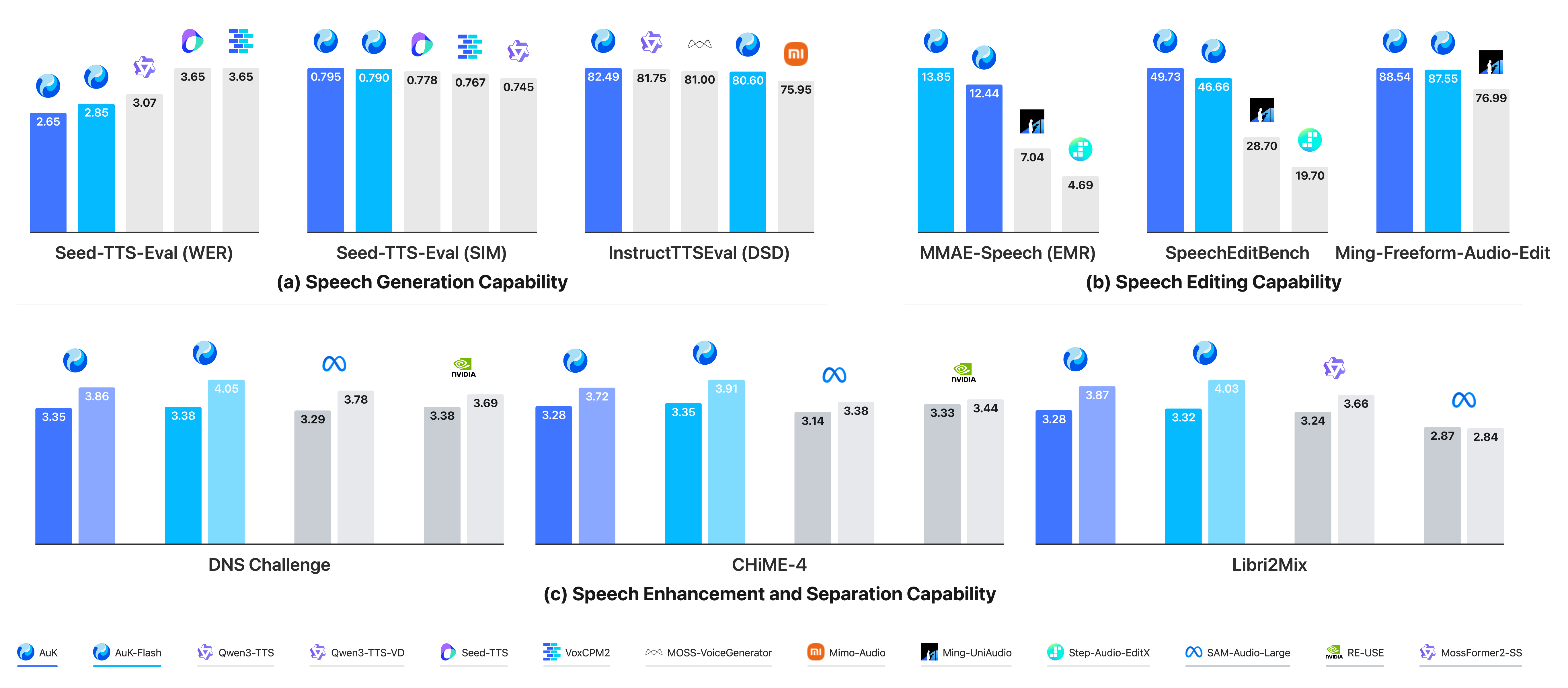}
\caption{\textbf{Performance comparison with SOTA models across speech generation, editing, enhancement and separation}. (a) Seed-TTS-Eval WER and SIM, averaged over test-en/test-zh/test-zh-hard, measure intelligibility and speaker similarity; InstructTTSEval DSD (mean of DSD-ZH/EN) measures how faithfully a model realises a free-form timbre instruction. (b) MMAE-Speech EMR measures exact-match success across all editing rubrics; SpeechEditBench averages five edit types, Ming-Freeform-Audio-Edit four semantic-editing splits. (c) Per model, DNSMOS-OVRL (darker, left) and UTMOS (lighter, right) rate perceptual quality; DNS Challenge and CHiME-4 are enhancement, Libri2Mix separation. Higher is better everywhere except WER, where shorter bars indicate fewer errors.}
\end{figure}

\clearpage
{
\fontsize{8.5}{8}\selectfont
\tableofcontents
}
\clearpage

\begin{figure*}[t]
    \centering
    \includegraphics[width=1.0\linewidth]{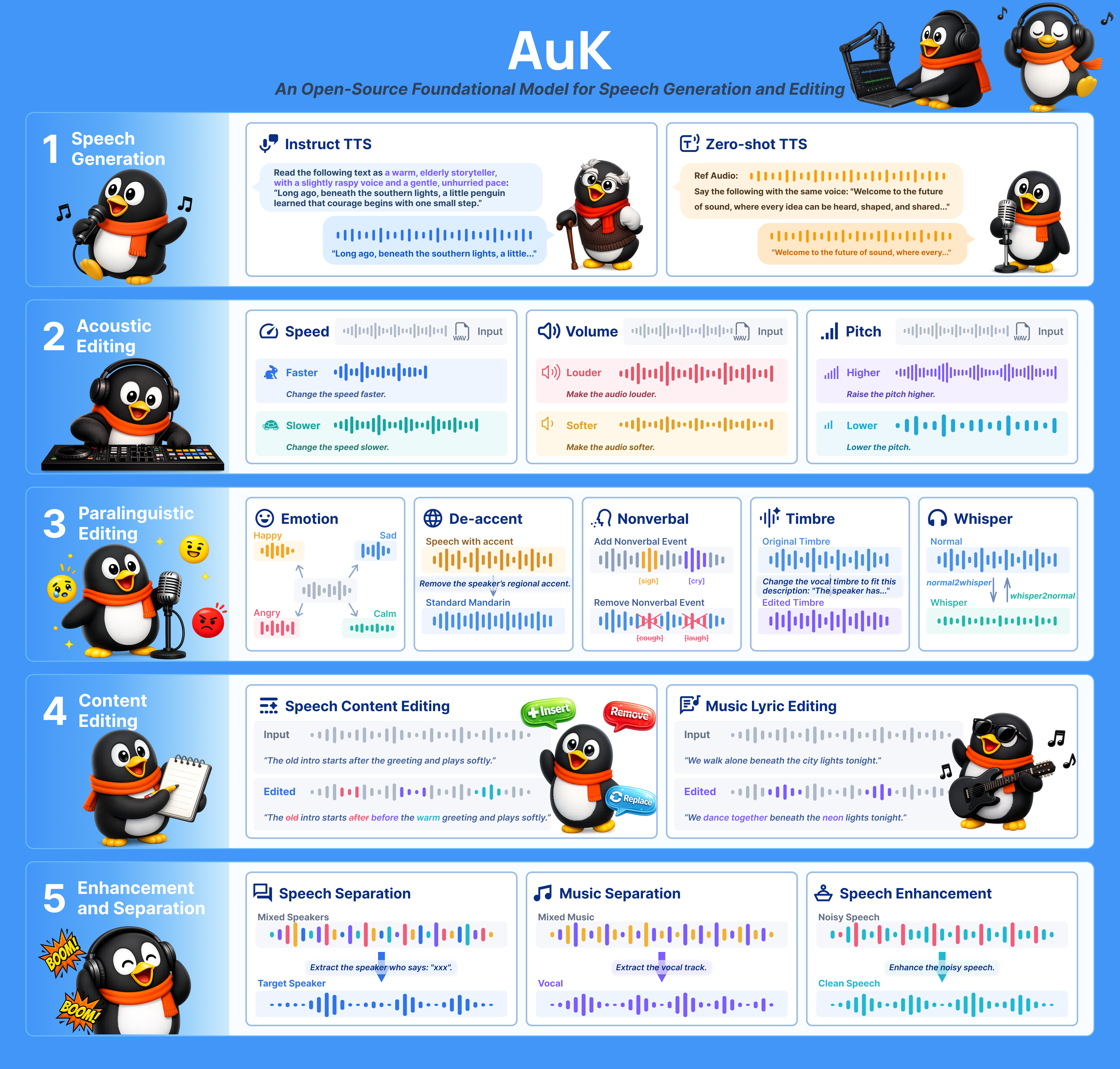}
    \caption{\textbf{Versatile speech generation and editing capabilities of \modelname.} The model supports five task families: (1) speech generation, including instruction-based and zero-shot TTS; (2) acoustic editing of speaking rate, loudness, and pitch; (3) paralinguistic editing of emotion, accent, nonverbal vocalizations, timbre, and whisper style; (4) content editing of spoken words and song lyrics; and (5) enhancement and separation of speech and music. The examples illustrate how natural-language instructions and speech input are mapped to output speech.}    \label{fig:capability-overview}
\end{figure*}

\section{Introduction}
Recent speech generation systems have advanced from conventional text-to-speech toward zero-shot voice cloning~\cite{cosyvoice, cosyvoice2, cosyvoice3, f5tts, xie2025fireredtts2, zhou2026indextts2, jiang2025megatts, seedtts, jia2025ditar, peng2025vibevoice, voxcpm2}, instruction-controlled synthesis~\cite{qwen3tts, mimoaudio, hu2026voicesculptor, zhou2024voxinstruct, guo2023prompttts, liu2023promptstyle, mossvoicegenerator}, and increasingly flexible speech editing~\cite{yan2025ming, yan2025step, mmedit, audio-omni, cosyedit, cosyedit2, lan2025guiding}. In practical use, however, these capabilities rarely appear in isolation. A user may ask a system to synthesize speech in a described style, replace part of an utterance, alter its emotion or accent, insert a nonverbal vocalization, isolate a speaker, or restore degraded audio. Supporting such requests with separate task-specific models fragments the user experience and duplicates modeling effort. This motivates a unified model that interprets free-form instructions and generates the requested speech.

Unifying these capabilities is challenging for three reasons. First, their output constraints differ fundamentally: generation creates new speech, content editing changes only selected regions, paralinguistic and acoustic editing must preserve linguistic content, and enhancement or separation must retain only the scene components specified by the instruction. Second, the conditioning interface varies across tasks. Some tasks rely on text alone, whereas others require joint reasoning over an instruction and source or reference audio. Third, supervision and evaluation are heterogeneous. Recognition accuracy and speaker similarity provide scalable signals for generation, but open-ended editing also depends on subjective judgments of naturalness, edit strength, contextual appropriateness, and preservation of unspecified attributes. Existing unified audio systems and editing benchmarks have begun to expose this broader problem~\cite{yan2025ming,yan2025step,ma2026mmae,zhang2026speecheditbench}, yet a single model with broad task coverage, scalable training, and efficient inference remains difficult to realize.

We introduce \modelname, a unified foundational model for speech generation and editing. As illustrated in \cref{fig:capability-overview}, \modelname spans speech generation, low-level acoustic control, paralinguistic transformation, content editing, and signal enhancement or separation. We formulate these capabilities through a common interface: a natural-language instruction and optional audio context are mapped to a target waveform. \modelname combines an MLLM semantic encoder, an audio VAE, and a hybrid flow Transformer. The MLLM~\cite{Qwen2.5-Omni} encodes either text alone or text jointly with reference audio, and aggregates hierarchical hidden states into a semantic condition. The \modelname-VAE, jointly trained on speech, general audio, and music, provides a shared acoustic latent space for reference conditioning and waveform reconstruction. Dual-stream MMDiT~\cite{flux-2-2025} blocks first exchange information between semantic and acoustic streams while preserving their distinct residual pathways; subsequent single-stream DiT blocks jointly refine the fused sequence and predict the rectified-flow velocity of the target latent. This design supports text-only generation and reference-conditioned editing within the same backbone.

Training is organized to address the different requirements of generation and editing. Unified pre-training begins with a generation-only warm-up and then jointly optimizes generation and editing tasks with a shared flow-matching objective. Post-training contains two complementary stages. For editing, where task completion is open-ended and no sufficiently broad preference dataset or reward model exists, we collect human feedback on free-form editing requests and perform flow-based preference optimization. For generation, we apply Flow-GRPO with automatic rewards for content correctness, speaker similarity, and instruction--style consistency. Finally, we distill the post-trained model through consistency initialization and task-routed Decoupled DMD. The resulting \modelname-Flash performs 4-step, CFG-free inference and achieves a $4.5\times$ wall-clock speedup over the $32$-NFE teacher under matched conditions.

Experiments cover \modelname-VAE reconstruction, zero-shot and instruction speech generation, general instruction-guided speech editing, speech enhancement, separation, and super-resolution. Across these evaluations, \modelname achieves leading performance on speech generation and general instruction-guided editing benchmarks while remaining competitive on signal-level restoration tasks. \modelname-Flash retains broad generation and editing capability under substantially reduced inference cost. We release both the source code and model weights to facilitate open-source community and further research.

\section{Data Construction}
\label{sec:data-pretraining}
\begin{figure*}[t]
    \centering
    \includegraphics[width=\textwidth]{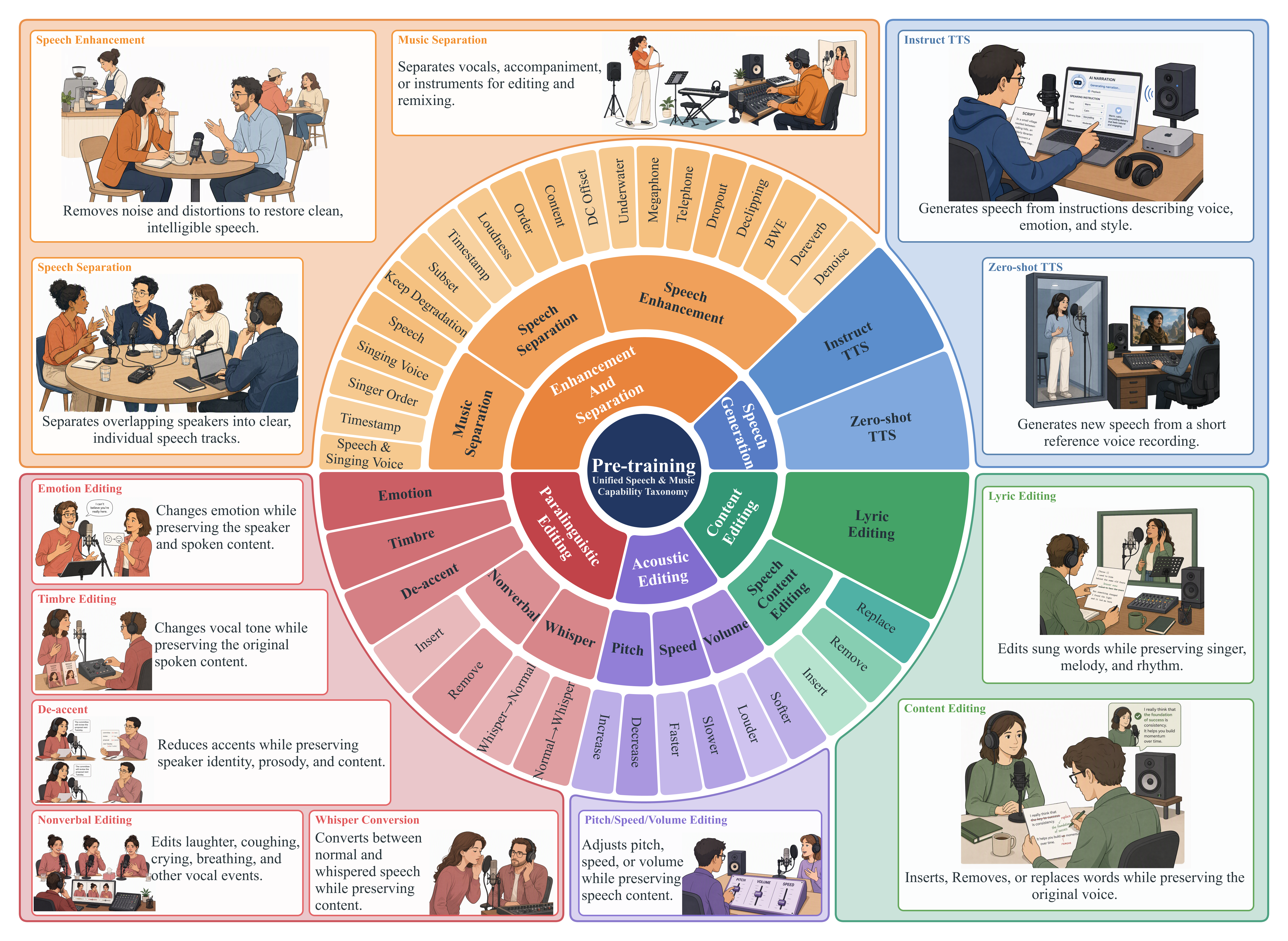}
    \caption{\textbf{Overview of the pre-training corpus.} The inner ring groups tasks into five capability families, while the outer rings summarize their instruction-level operations. The surrounding panels illustrate representative tasks and their intended functions. Sector widths are adjusted for readability and do not indicate data volume.}
    \label{fig:pretrain-data-taxonomy}
\end{figure*}

As summarized in \cref{fig:pretrain-data-taxonomy}, we organize the pre-training corpus into five task families: \textbf{speech generation, acoustic editing, paralinguistic editing, content editing, and enhancement and separation}. Despite their different objectives, all tasks share a unified interface consisting of a natural-language instruction, optional input audio, and a target waveform. This formulation allows a single model to learn generation, restoration, separation, and editing from approximately $3.03$ billion instruction--audio instances, with a total of $1.95$ million hours of effective audio supervision.

\subsection{Speech Generation}
Speech generation teaches the model to synthesize natural, intelligible, and controllable speech from text. We construct two complementary forms of supervision: transcript-free zero-shot TTS conditioned on reference speech, and instruct TTS controlled by free-form descriptions.

\paragraph{Zero-Shot TTS.}
We build a large-scale bilingual corpus through a multi-stage curation pipeline. Source separation and speech enhancement are first applied to improve signal quality, followed by MOS-based quality filtering, speaker-identity verification, and cross-validation with multiple ASR systems to remove noisy or inconsistent utterances.

Conventional zero-shot TTS often requires both a reference utterance and its transcript, making deployment dependent on an additional ASR system. We instead formulate zero-shot TTS as transcript-free in-context learning. For a speaker with $n$ distinct utterances, we enumerate all $C_n^2$ unordered pairs and assign each utterance in a pair once as the acoustic prompt and once as the synthesis target, yielding $n \times (n-1)$ bidirectional training instances. Each instance contains only the prompt speech and target text; the prompt transcript is never provided. At inference time, the model can therefore clone a speaker from either a complete reference utterance or a randomly cropped segment without requiring a transcription.

\paragraph{Instruct TTS.}
We derive the Instruct TTS corpus from a large-scale bilingual speech collection that has undergone standardized preprocessing and quality control. Qwen3-Omni~\cite{Qwen3-Omni} annotates each retained utterance with a free-form natural-language caption and structured attributes covering gender, age, speaking rate, clarity, fluency, vocal state, intonation, loudness, timbre, pitch, accent, emotion, and personality. We combine this description with the target text to form the input instruction and use the corresponding waveform as the synthesis target. Because no reference audio is provided, the resulting pairs teach the model to design the voice directly from the expressive descriptions.
\subsection{Acoustic Editing}
Acoustic editing teaches the model to control low-level speaking attributes, including speaking rate, pitch, and loudness, while preserving linguistic content, speaker identity, and all non-target characteristics. We generate paired supervision using deterministic signal-processing transformations.

For each source utterance, we create targets at five speaking-rate multipliers ($0.5\times$, $0.75\times$, $1.25\times$, $1.5\times$, and $2.0\times$), six loudness offsets ($\pm5$, $\pm10$, and $\pm15$\,dB), and six pitch shifts ($\pm1$, $\pm2$, and $\pm3$ semitones). Speaking rate is modified using pitch-preserving time stretching, pitch using duration-preserving pitch shifting, and loudness using waveform gain. Each transformed waveform is paired with its source and a natural-language instruction specifying the attribute and requested magnitude. We validate every transformation and apply peak protection when necessary to prevent clipping. 
\subsection{Paralinguistic Editing}
Paralinguistic editing teaches the model to modify how an utterance is delivered while preserving what is said. We construct paired supervision for emotion, timbre, accent, nonverbal vocalization, and whisper-style editing.

\paragraph{Emotion Editing.}
We construct emotion editing samples from the bilingual speech pool used for Instruct TTS to ensure that the speech is expressive. Target emotions are sampled from eight categories: angry, happy, sad, fearful, surprised, disgusted, calm, and excited. Given a source transcript and target-emotion instruction, Qwen3-TTS-CustomVoice~\cite{qwen3tts} first synthesizes an expressive reference utterance. IndexTTS2~\cite{zhou2026indextts2} is then conditioned on the original utterance for speaker characteristics and on the synthesized reference for emotion, while the transcript remains fixed. The resulting waveform is paired with the source audio and a natural-language editing instruction. 

\paragraph{Timbre Editing.}
We adopt the X-VC~\cite{x-vc} training corpus, which is constructed using SeedVC-Small~\cite{seedvc}. Each group contains four aligned source--target pairs that preserve linguistic content while changing speaker timbre. All waveforms are enhanced with speech super-resolution and standardized to a sampling rate of $24$\,kHz. Qwen3-Omni~\cite{Qwen3-Omni} generates a natural-language timbre description for each target waveform, which is incorporated into the corresponding editing instruction. 

\paragraph{De-accent.}
We construct de-accenting pairs from an in-house corpus spanning $13$ Chinese dialect and regional-accent categories. For each accented source utterance, CosyVoice2~\cite{cosyvoice2} first synthesizes a same-speaker standard Mandarin reference from independently sampled text, using the source utterance as the speaker prompt. We then partially mask the source and use OmniVoice~\cite{zhu2026omnivoice} to reconstruct it conditioned on the source transcript and synthesized standard Mandarin reference. The target follows standard Mandarin pronunciation while preserving the source speaker's timbre and prosodic characteristics. 

\paragraph{Nonverbal Editing.}
We build the nonverbal-editing corpus from both public datasets and in-house datasets with heterogeneous human- and model-derived annotations. We normalize these annotations into $39$ event types spanning physiological sounds, affective expressions, and discourse vocalizations. For each event, Qwen3-ForcedAligner~\cite{qwen3-asr} locates its temporal span. We mask the event and its immediate context, then use F5-TTS~\cite{f5tts} to reconstruct an event-free waveform while preserving the surrounding speech, speaker identity, and prosody. Each original--reconstructed pair supports both event removal and insertion, with a natural-language instruction specifying the event type and location. 

\paragraph{Whisper-Style Conversion.}
We construct normal-to-whisper pairs from public Mandarin corpus containing parallel normal and whispered speech. We retain only pairs whose normal and whispered transcripts satisfy $\mathrm{WER}=0$. All recordings are resampled to $24$\,kHz, and the normal-speech inputs are normalized per utterance to an RMS target of $-24$\,dBFS, matching the average level of the broader speech training corpus. 

\subsection{Content Editing}
Content editing teaches the model to insert, delete, or replace spoken and sung content while preserving speaker identity, prosody, melody, and the acoustic context outside the edited region. We construct paired supervision for both speech-content and lyric-content editing.

\paragraph{Speech Content Editing.}
Starting from high-quality transcribed speech, we use a large language model (LLM) to generate operator-specific annotations and target transcripts for insertion, deletion, and substitution. Each operation is applied independently, yielding examples with an explicit edit type and a well-defined target transcript.

We synthesize the target waveform through localized masked infilling. Qwen3-ForcedAligner~\cite{qwen3-asr} first provides word-level alignments between the source waveform and transcript, allowing each edited span to be mapped to its temporal interval. We mask only these intervals and condition F5-TTS~\cite{f5tts} on the masked source waveform and target transcript to generate the requested content. This construction modifies only the designated region while preserving speaker identity, prosody, and acoustic context elsewhere.
We transcribe each synthesized waveform and compare it with the target transcript for quality control, and retain only samples with low word error rate. 

\paragraph{Lyric Editing.}
We construct lyric-editing data from high-quality dry-vocal recordings. Each recording is transcribed and aligned to its lyrics at the word level, after which an LLM generates source--target lyric pairs for localized edits. Chinese replacements preserve the number of characters and are checked at the pinyin level, whereas English replacements respect complete word boundaries and preserve the number of words.

YingMusic-Singer-Plus~\cite{hao2026yingmusic} then synthesizes the target vocal. We mask only the latent interval associated with the edited lyrics and condition generation on the complete target lyrics and original melody, preserving the singer's timbre, rhythm, expression, and surrounding acoustic details. As in speech-content editing, we transcribe each synthesized vocal and retain only samples with low word error rate. 

\subsection{Enhancement and Separation}
Enhancement and separation train the model to transform a complex acoustic scene according to a natural-language request. Given the same mixture, the instruction determines which components should be preserved, removed, isolated, or restored, providing unified supervision for speech enhancement, source extraction, source removal, and selective editing.

\paragraph{Speech Enhancement.}
We construct speech-enhancement examples by applying independently sampled degradations to quality-filtered speech. Candidate non-speech recordings are first transcribed with an ASR system, and clips containing intelligible words are discarded before the remaining audio is mixed as sustained background noise or localized acoustic events. Reverberation is introduced using both measured room impulse responses and simulated rooms with randomized geometry, reverberation time, source locations, and microphone locations. We additionally apply channel degradations, including bandwidth limitation, clipping, signal dropout, telephone and megaphone coloration, underwater-like filtering, and DC offset. Randomizing the type, severity, and combination of these degradations prevents the model from associating an instruction with a single acoustic signature.

Training targets are not limited to fully clean speech. In addition to recovering the original signal, we create selective targets that remove only the corruption named by the instruction. For example, a denoising target may retain reverberation, a dereverberation target may retain environmental sound, and a channel-restoration target may preserve all unrelated scene attributes. The model therefore learns that enhancement is instruction-dependent: a component removed for one request may be intentionally retained for another.

\paragraph{Multi-Speaker Separation.}
We construct conversational mixtures by arranging multiple speakers on a shared timeline with turn-taking, pauses, interruptions, and partial or complete overlap. Speaker gain, temporal placement, room acoustics, and optional background interference are varied independently, producing scenes that better resemble natural conversations than simple waveform addition. 

Natural-language instructions identify the desired speakers through complementary cues, including spoken content, speaking order, relative loudness, or an exclusive timestamp. An instruction may retain or remove one speaker or a subset of speakers. We also construct targets that remove a speaker while preserving the scene's noise, reverberation, and channel effects. Solving these examples requires the model to interpret the request, locate the relevant source, and preserve all components that are not explicitly targeted.

\paragraph{Music Enhancement and Separation.}
For music data, we first obtain aligned vocal and accompaniment stems using a source-separation model. The accompaniment is transcribed and compared with the vocal transcript or lyrics; songs with substantial textual overlap are rejected to reduce residual singing in the accompaniment stem. We then construct two complementary forms of supervision. Native-song examples use the original song as input and time-aligned stem crops as targets, avoiding artifacts that would arise from reconstructing the input from separated tracks. Scene-based examples combine speech, singing voices, and background music with controlled timing and gain, creating mixtures such as speech over music, speech mixed with singing, and multiple overlapping singers.

The resulting instructions request spoken speech, a singing voice, a group of vocal sources, or a singer identified by order or timestamp. This design casts music processing as the same instruction-conditioned source-selection problem used for conversational separation while retaining the acoustic complexity of real songs.

\section{Model Design}
\begin{figure*}[t]
    \centering
    \makebox[\textwidth][c]{
        \includegraphics[width=1.\textwidth]{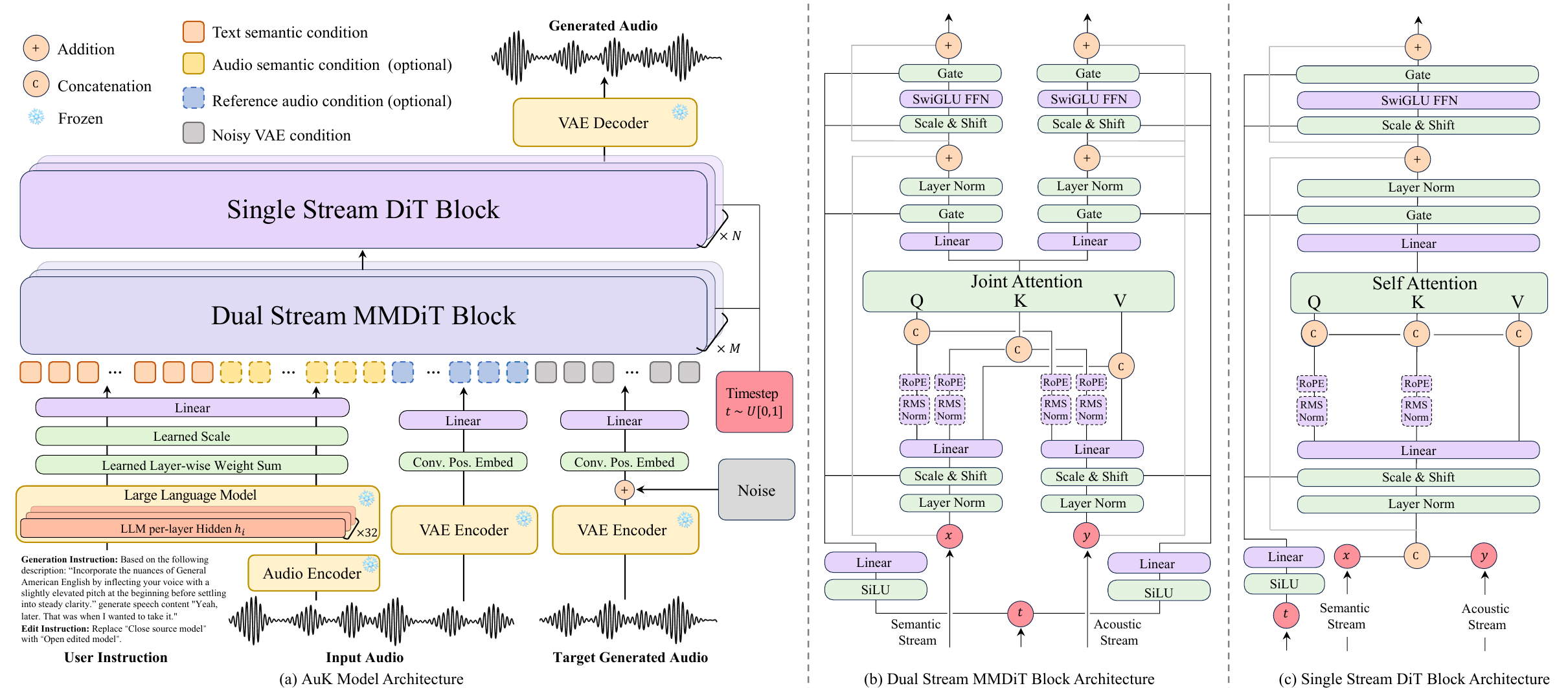}
    }
    \caption{\textbf{Architecture of \modelname.} (a) The framework maps a user instruction and optional input audio to complementary semantic and acoustic conditioning streams. A multimodal language model encodes the instruction and audio context, and a learnable weighted sum of its layer-wise hidden states forms the semantic condition. In parallel, the VAE maps input audio, when present, to reference latents, which are combined with noisy target latents to form the acoustic condition. The two streams exchange information through $M$ dual-stream MMDiT blocks before being concatenated and refined by $N$ single-stream DiT blocks. The predicted latent is decoded by the VAE to produce the output audio. (b) Each MMDiT block performs joint attention over semantic and acoustic tokens while preserving stream-specific residual pathways. (c) Each DiT block applies self-attention to the fused token sequence. The symbols $+$ and {\scshape c} denote addition and concatenation, respectively; snowflakes denote frozen modules.}
    \label{fig:model-overview}
\end{figure*}

\subsection{Overall Architecture}
Figure~\ref{fig:model-overview} shows the complete architecture of \modelname. It consists of three components with complementary roles:
the MLLM jointly processes textual instruction and audio context to produce the multimodal semantic condition; the pre-trained Variational Autoencoder (VAE) encodes audio into a latent space that preserves fine-grained acoustic information; and a FLUX-style\cite{flux-2-2025} transformer backbone promotes interaction and fusion between the semantic and acoustic conditions while predicting the target acoustic latent.

During training, we group tasks into two categories based on input audio availability. \textbf{(1) Tasks with reference audio} (e.g., zero-shot TTS, content editing, speech enhancement and separation). The input audio is sent to both the audio encoder of an MLLM and the VAE encoder. The audio encoder supplies audio representations to the LLM, while the textual user instruction is provided directly to the LLM. In parallel, the VAE encoder converts the same input audio into reference acoustic latents. The semantic condition, reference acoustic latents and noisy target latents are concatenated along the
sequence dimension.
\textbf{(2) Tasks without reference audio} (e.g., Instruct TTS). The user instruction is processed directly by the LLM; no audio is sent to the audio encoder or the reference branch of the VAE encoder. In this case, the acoustic stream contains only the noisy target latents. After $M$ dual-stream blocks, the semantic and acoustic streams are concatenated along the sequence dimension and refined by $N$ single-stream blocks. In both configurations, the hybrid transformer predicts the denoised latent, which is converted to the audio by the VAE decoder.

\subsection{MLLM Semantic Condition}
A fixed output layer of a multimodal language model may not provide optimal conditioning across diverse audio generation and editing tasks, since representations at different depths capture complementary linguistic, acoustic, and cross-modal cues. To retain this information, \modelname uses Qwen2.5-Omni~\cite{Qwen2.5-Omni} as its semantic encoder and aggregates its layer-wise hidden states.

Let $\mathbf{t}$ denote the tokenized user instruction and $\mathcal{E}_{\mathrm{aud}}(\mathbf{x}_{\mathrm{ref}})$ denote the audio encoder extracted from an optional reference waveform. The hidden state at MLLM layer $\ell$ is
\begin{equation}
    \mathbf{h}^{(\ell)} =
    \begin{cases}
        \operatorname{MLLM}^{(\ell)}
        \!\left(\mathbf{t},\mathcal{E}_{\mathrm{aud}}(\mathbf{x}_{\mathrm{ref}})\right),
        & \text{with reference audio},\\
        \operatorname{MLLM}^{(\ell)}(\mathbf{t}),
        & \text{otherwise}.
    \end{cases}
    \label{eq:mllm-semantic-input}
\end{equation}
Thus, audio-conditioned tasks encode the instruction and reference audio jointly, whereas text-only tasks derive their semantic representation solely from the instruction.

The layer-wise representations are aggregated into the semantic condition as
\begin{equation}
    \mathbf{c}_{\mathrm{sem}}
    = \sum_{\ell=1}^{L} w_{\ell} \cdot
      \operatorname{LayerNorm}\!\left(\mathbf{h}^{(\ell)}\right),
    \label{eq:semantic-layer-fusion}
\end{equation}
where $L$ is the number of MLLM layers and $w_{\ell}$ is an unconstrained learnable scalar for layer $\ell$. Layer normalization~\cite{layernorm} balances the scale of representations across layers before aggregation. The resulting $\mathbf{c}_{\mathrm{sem}}$ combines information from multiple levels of abstraction and is used as the semantic condition of the generative backbone.

\subsection{VAE Acoustic Condition and Reconstruction}
We employ a flow-augmented audio VAE~\cite{zhang2025minimaxspeech} to provide a shared latent space for reference-audio conditioning and waveform reconstruction. Its non-causal encoder $\mathcal{E}_{\mathrm{enc}}$ maps a $24$\,kHz waveform to a sequence of $64$-dimensional latents at $50$\,Hz, while its causal decoder $\mathcal{E}_{\mathrm{dec}}$ maps a latent sequence back to audio. A normalizing flow regularizes the latent distribution during VAE training.

For a reference waveform $\mathbf{x}_{\mathrm{ref}}$, the encoder produces posterior parameters and a latent sample
\begin{equation}
    (\boldsymbol{\mu}_{\mathrm{ref}},\log\boldsymbol{\sigma}_{\mathrm{ref}})
    = \mathcal{E}_{\mathrm{enc}}(\mathbf{x}_{\mathrm{ref}}),
    \qquad
    \mathbf{z}_{\mathrm{ref}}
    = \boldsymbol{\mu}_{\mathrm{ref}}
    + \boldsymbol{\epsilon}\odot\boldsymbol{\sigma}_{\mathrm{ref}},
    \quad
    \boldsymbol{\epsilon}\sim\mathcal{N}(\mathbf{0},\mathbf{I}).
\end{equation}
The acoustic condition is the VAE-encoder latent itself:
\begin{equation}
    \mathbf{c}_{\mathrm{ac}} =
    \begin{cases}
        \mathbf{z}_{\mathrm{ref}},
        & \text{if reference audio is available},\\
        \varnothing,
        & \text{otherwise}.
    \end{cases}
    \label{eq:acoustic-condition}
\end{equation}
Here, $\mathbf{c}_{\mathrm{ac}}$ is the unprojected reference latent in the VAE space and is projected into the Transformer hidden dimension only when passed to the backbone, as described in Eq.~\eqref{eq:condition-mapping}. During VAE training, the encoder--decoder pair is optimized to reconstruct the input waveform. During generation, the causal decoder $\mathcal{E}_{\mathrm{dec}}$ converts the resulting target latent into the output waveform. Detailed descriptions of the VAE architecture and training objective are provided in Section~\ref{sec:vae-training}.

\subsection{Transformer Backbone}
The Transformer backbone of \modelname follows a FLUX-style~\cite{flux-2-2025} hybrid design. Its first $M$ layers are dual-stream MMDiT blocks~\cite{esser2024scaling}, and the following $N$ layers are single-stream DiT blocks~\cite{Peebles2022DiT}. A sinusoidal embedding of the flow time $t\sim\mathcal{U}[0,1]$ is projected by an MLP and used to modulate every block.

\paragraph{Condition Mapping.}
The semantic condition, optional acoustic condition, and noisy target latent are mapped to the two input streams as
\begin{equation}
    \mathbf{s}^{(0)}
    = \mathcal{P}_{\mathrm{sem}}(\mathbf{c}_{\mathrm{sem}}),
    \qquad
    \mathbf{a}^{(0)}
    = \left[
        \mathcal{P}_{\mathrm{ref}}(\mathbf{c}_{\mathrm{ac}});
        \mathcal{P}_{\mathrm{tgt}}(\mathbf{z}_t)
      \right].
    \label{eq:condition-mapping}
\end{equation}
Here, $\mathbf{c}_{\mathrm{sem}}$ is the MLLM semantic condition, $\mathbf{c}_{\mathrm{ac}}$ is the optional VAE reference latent, and $\mathbf{z}_t$ is the noisy target latent at flow time $t$. The operator $\mathcal{P}_{\mathrm{sem}}$ consists of a linear projection
and RMSNorm, whereas $\mathcal{P}_{\mathrm{ref}}$ and $\mathcal{P}_{\mathrm{tgt}}$ each consist of the convolutional positional embedding and linear projection shown in Figure~\ref{fig:model-overview}. 

\paragraph{Dual-Stream MMDiT Block.}
The first $M$ blocks jointly update the semantic and acoustic streams:
\begin{equation}
    \left(\mathbf{s}^{(M)},\mathbf{a}^{(M)}\right)
    = \operatorname{MMDiT}^{M}
      \left(\mathbf{s}^{(0)},\mathbf{a}^{(0)};\mathbf{e}_t\right).
    \label{eq:mmdit-denoising}
\end{equation}
Here, $\mathbf{e}_t$ is the projected flow-time embedding, $\operatorname{MMDiT}^{M}$ denotes the stack of $M$ MMDiT blocks, and $\mathbf{s}^{(M)}$ and $\mathbf{a}^{(M)}$ are its semantic and acoustic outputs. Each block uses stream-specific query, key, value, and residual projections. RoPE~\cite{su2024roformer} is applied according to the positions of each stream before their attention tensors are concatenated for joint attention, enabling bidirectional semantic--acoustic interaction while preserving separate residual pathways.

\paragraph{Single-Stream DiT Block.}
The updated streams are concatenated and processed by the subsequent $N$ DiT blocks to predict the flow velocity:
\begin{equation}
    \widehat{\mathbf{v}}_t
    = \mathcal{P}_{\mathrm{out}}
      \left(
        \operatorname{DiT}^{N}
        \left(
          [\mathbf{s}^{(M)};\mathbf{a}^{(M)}];\mathbf{e}_t
        \right)_{\mathrm{tgt}}
      \right).
    \label{eq:dit-denoising}
\end{equation}
Here, $\operatorname{DiT}^{N}$ denotes the stack of $N$ DiT blocks, the subscript $\mathrm{tgt}$ selects the target-latent positions from its output, $\mathcal{P}_{\mathrm{out}}$ is the final output projection, and $\widehat{\mathbf{v}}_t$ is the predicted flow velocity. 
Both MMDiT and DiT blocks apply RMSNorm-based~\cite{rmsnorm} QK-Norm before RoPE and use zero-initialized, time-conditioned AdaLN~\cite{Peebles2022DiT} to modulate their attention and SwiGLU feed-forward networks~\cite{shazeer2020glu}.

\section{Model Training}
\subsection{VAE Training}
\label{sec:vae-training}
The \modelname-VAE is trained to provide a compact latent space that supports both reference-audio conditioning and high-fidelity waveform reconstruction. We describe its architecture, training data, and optimization objective below.

\paragraph{Architecture Configuration.}
The \modelname-VAE operates on $24$\,kHz mono waveforms. The encoder first projects the waveform to $12$ channels with a convolution of kernel size $3$, followed by six downsampling blocks with strides $(2,2,2,3,4,5)$ and channel widths $12\rightarrow24\rightarrow48\rightarrow96\rightarrow192\rightarrow384\rightarrow768$. Each block contains a strided convolution whose kernel size is twice its stride, six residual units with kernel size $3$ and dilation rates $(1,2,4,8,16,32)$, and a final LeakyReLU activation. A final convolution with kernel size $3$ produces $128$ channels, which parameterize a $64$-dimensional posterior mean and log standard deviation. The overall downsampling factor is $480$, yielding a latent frame rate of $50$\,Hz. All encoder convolutions use weight normalization.

A normalizing flow $\mathcal{F}$, composed of four residual coupling layers interleaved with channel flips, maps the posterior sample $\mathbf{z}$ to $\mathbf{z}_p=\mathcal{F}(\mathbf{z})$ for latent regularization. The flow is used only during VAE training; acoustic conditioning and waveform reconstruction operate in the original latent space $\mathbf{z}$.

The decoder follows BigVGAN~\cite{lee2022bigvgan}. A $3$-frame look-ahead convolution with kernel size $7$ projects the latent representation to $1536$ channels, after which all convolutions are causal. Six transposed-convolution blocks use strides $(5,4,3,2,2,2)$ and kernels twice their strides while reducing the channel width as $1536\rightarrow768\rightarrow384\rightarrow192\rightarrow96\rightarrow48\rightarrow24$. Each block is followed by an anti-aliased multi-periodicity composition module with residual branches of kernel sizes $\{3,7,11\}$ and dilation rates $\{1,3,5\}$. The decoder uses channel-wise SnakeBeta activations and a final causal convolution with kernel size $7$ to reconstruct the waveform.

\paragraph{Training Data and Setup.}
We train the \modelname-VAE for $1.24$ million updates on approximately $3$ million hours of speech, music, and general audio. The three domains are sampled at the instance level with a ratio of $6{:}3{:}1$. All waveforms are resampled to $24$\,kHz and divided into $1.28$-second segments.

\paragraph{Training Objectives.}
Following DAC~\cite{kumar2023high}, we use a multi-scale log-mel reconstruction loss $\mathcal{L}_{\mathrm{mel}}$, defined as the sum of $\ell_1$ distances between the mel spectrograms of $\mathbf{x}$ and $\widehat{\mathbf{x}}$ at multiple resolutions. We use window lengths $\{32,64,128,256,512,1024,2048\}$, hop sizes $\{8,16,32,64,128,256,512\}$, and mel-bin counts $\{5,10,20,40,80,160,320\}$ to capture both short-time acoustic detail and long-range spectral structure.

To improve perceptual quality, we apply adversarial training with a multi-period discriminator~\cite{kong2020hifi} using periods $\{2,3,5,7,11\}$ and a multi-scale CQT discriminator~\cite{gu2023multiscalesubbandconstantqtransform}. Both discriminator families use the least-squares GAN objective~\cite{mao2017least}; an additional $\ell_1$ feature-matching loss $\mathcal{L}_{\mathrm{feat}}$ is computed from their intermediate representations. We regularize the transformed posterior $q_\phi(\mathbf{z}_p\mid\mathbf{x})$ toward a standard Gaussian:
\begin{equation}
    \mathcal{L}_{\mathrm{KL}}
    = D_{\mathrm{KL}}\!\left(
        q_\phi(\mathbf{z}_p\mid\mathbf{x})
        \middle\|
        \mathcal{N}(\mathbf{0},\mathbf{I})
      \right).
\end{equation}
The complete generator objective is
\begin{equation}
    \mathcal{L}_{\mathrm{VAE}}
    = \lambda_{\mathrm{mel}}\mathcal{L}_{\mathrm{mel}}
    + \lambda_{\mathrm{adv}}\mathcal{L}_{\mathrm{adv}}
    + \lambda_{\mathrm{feat}}\mathcal{L}_{\mathrm{feat}}
    + \lambda_{\mathrm{KL}}\mathcal{L}_{\mathrm{KL}},
    \label{eq:vae-training-objective}
\end{equation}
where $\mathcal{L}_{\mathrm{adv}}$ is the generator-side least-squares GAN loss. We set $\lambda_{\mathrm{mel}}=15$, $\lambda_{\mathrm{adv}}=1$, $\lambda_{\mathrm{feat}}=2$, and $\lambda_{\mathrm{KL}}=5$. The discriminators are optimized separately with their corresponding discriminator objective.

\subsection{Unified Pre-Training}
Unified pre-training follows a two-stage curriculum. The first stage establishes basic speech-generation capability using only generation tasks, while the second stage jointly optimizes generation and editing with a shared rectified-flow objective, condition-dropout strategy, and task mixture. Throughout both stages, the MLLM semantic encoder and audio VAE remain frozen; only the Transformer backbone and layer-fusion parameters are updated.

\paragraph{Model Configuration.}
The backbone contains $30$ Transformer layers: $10$ dual-stream MMDiT blocks followed by $20$ single-stream DiT blocks. Every block has a hidden dimension of $1536$, $24$ attention heads with a head dimension of $64$, and a SwiGLU feed-forward network with an intermediate dimension of $3072$. The backbone contains approximately $1.5$ billion parameters and adopts the RoPE and convolutional positional embedding design used in F5-TTS~\cite{f5tts}.

\paragraph{Training Curriculum.}
In the first stage, we train exclusively on speech-generation tasks for $50$k updates. This generation-only warm-start establishes stable text-to-speech alignment and basic synthesis quality before the model is exposed to heterogeneous editing objectives. The second stage initializes from this checkpoint and jointly trains on generation and editing tasks for a further $600$k updates. The five-family sampling mixture reported below applies to this second stage.\paragraph{Flow-Matching Objective.}
Let $\mathbf{z}_1$ denote the clean target latent produced by the frozen VAE encoder and let $\mathbf{z}_0\sim\mathcal{N}(\mathbf{0},\mathbf{I})$ denote Gaussian noise. For a sampled flow time $t\in[0,1]$, we construct the linear interpolation and target velocity as
\begin{equation}
    \mathbf{z}_t=(1-t)\mathbf{z}_0+t\mathbf{z}_1,
    \qquad
    \mathbf{v}_t=\mathbf{z}_1-\mathbf{z}_0.
    \label{eq:flow-matching-path}
\end{equation}
The model predicts $\widehat{\mathbf{v}}_t$ from $\mathbf{z}_t$, the flow-time embedding, and the available semantic and acoustic conditions. Training minimizes a masked mean-squared error over valid, non-padding latent frames:
\begin{equation}
    \mathcal{L}_{\mathrm{FM}}
    = \frac{
        \left\|
          \mathbf{m}\odot
          (\widehat{\mathbf{v}}_t-\mathbf{v}_t)
        \right\|_2^2
      }{
        \sum_i m_i
      },
    \label{eq:flow-matching-loss}
\end{equation}
where $\mathbf{m}$ is the validity mask. We sample $t=\sigma(u)$ with $u\sim\mathcal{N}(-0.8,0.8^2)$, biasing training toward smaller $t$, corresponding to noisier states closer to the Gaussian-noise endpoint, while retaining support over the full trajectory. At inference, the learned velocity field is integrated with a deterministic ODE solver.

\paragraph{Condition Dropout and Reference Augmentation.}
To support classifier-free guidance for both text-only generation and reference-conditioned tasks, we apply hierarchical condition dropout. We first sample an acoustic-condition dropout mask with probability $0.3$. We then independently sample an unconditional mask with probability $0.2$; when active, this second mask overrides the first decision and drops both the acoustic and semantic conditions. The resulting training examples cover fully conditioned, text-only, and unconditional configurations under the same objective.

Reference audio at inference can vary substantially in duration for tasks like zero-shot TTS. To reduce sensitivity to the reference lengths observed during training, with probability $0.5$ we crop zero-shot TTS references to a uniformly sampled duration between $3$ seconds and its original length.

\paragraph{Task Mixture and Dynamic Batching.}
During the second stage, we realize the five task families in \cref{sec:data-pretraining} as fixed per-batch sampling probabilities, summarized in \cref{tab:pretrain_family_mix}. Utterances range from $1$ to $35$ seconds and are grouped with dynamic length-bucketed batching. Each accelerator processes at most $10{,}000$ latent frames at $50$\,Hz or $24$ utterances per micro-batch, whichever limit is reached first. The resulting global batch contains at most $6{,}144$ utterances, corresponding to approximately $14$ hours of audio per optimizer step across GPUs.

\begin{table}[t]
\centering
\small
\setlength{\tabcolsep}{6pt}
\renewcommand{\arraystretch}{1.1}
\caption{Fixed per-batch sampling probabilities used in the second pre-training stage for the five task families introduced in \cref{sec:data-pretraining}.}
\label{tab:pretrain_family_mix}
\begin{tabular}{lr}
\toprule
\textbf{Task family} & \textbf{Sampling probability} \\
\midrule
Speech Generation & 28.10\% \\
Content Editing & 23.02\% \\
Enhancement and Separation & 23.17\% \\
Paralinguistic Editing & 21.75\% \\
Acoustic Editing & 3.96\% \\
\bottomrule
\end{tabular}
\end{table}

\paragraph{Optimization.}
We train on $256$ GPUs with DeepSpeed ZeRO-2 and bf16 precision, without CPU offload. Gradients are clipped to a global norm of $1.0$. We use fused AdamW with $\beta=(0.9,0.95)$ and a peak learning rate of $1\times10^{-4}$, linearly warmed up over the first $2{,}000$ updates and then held constant. An exponential moving average of the model weights with decay $0.9999$ is enabled after $100$ updates and refreshed every $10$ updates; the EMA checkpoint is used for evaluation and release.

\subsection{Post-Training}
\label{ssec:posttraining}
Post-training consists of two complementary stages. First, \textbf{editing preference optimization} uses human feedback to align diverse editing behaviors with subjective judgments of task completion and perceptual quality. Second, \textbf{generation reinforcement learning} uses automatic reward models to improve zero-shot and instruction-following speech generation. Both stages update only the Transformer backbone; the Qwen2.5-Omni encoder and audio VAE remain frozen.

\subsubsection{Editing Preference Optimization}
\label{sssec:editing_rl}
\paragraph{Human Preference Data.}
Editing tasks are judged by more subjective criteria than generation tasks. Whether a result is acceptable often depends on hard-to-quantify judgments of naturalness, degree of transformation, and contextual appropriateness, unlike the content correctness and speaker similarity in speech generation, which can be reliably measured by a single automatic metric. Moreover, for the diverse set of editing tasks we cover, neither a sufficiently broad preference dataset nor a ready-made reward model exists. We therefore collect human feedback directly and optimize editing behavior with preference learning.

For each editing instruction, we sample $10$ or $20$ candidate outputs from the pre-trained model. Annotators rate each candidate on a three-level ordinal scale $s_i\in\{0,1,2\}$, corresponding to failure, partial completion, and successful completion. Groups in which all $20$ candidates receive the same rating are discarded because they provide no preference signal. After filtering the dataset contains $818$ informative groups and $9{,}080$ rated candidates.

\paragraph{Flow-Based DPO Score.}
Following Diffusion-DPO~\cite{wallace2024diffusiondpo}, we define an implicit preference score from the improvement of the policy over a frozen reference model in flow-matching error. Although candidate waveforms are generated from independent noise trajectories, the DPO loss reevaluates candidates within a group using a shared noise sample $\mathbf{z}_0$ and flow time $t$, isolating differences attributable to the candidate target latents. The frozen-reference prediction is cached within each micro-batch.
\paragraph{Ordinal Listwise Objective.} We optimize the ordinal feedback with the listwise formulation of LiPO~\cite{liu2024lipo}, where $R_i^{\mathrm{pref}}$ is the implicit preference score defined above. Let $\mathcal{I}_q^+$ and $\mathcal{I}_q^-$ denote the higher- and lower-rated candidate sets for relation $q\in\{1\succ0,2\succ0,2\succ1\}$, and let $\Delta R_{ij}=R_i^{\mathrm{pref}}-R_j^{\mathrm{pref}}$ for a pair $(i,j)$. Each preference pair is penalized by a label-smoothed logistic loss ($\epsilon=0.05$), reduced in two levels: first averaged within each relation, then combined across relations with confidence weights:
\begin{equation}
\mathcal{L}_q= \frac{1}{|\mathcal{I}_q^+||\mathcal{I}_q^-|}\sum_{i\in\mathcal{I}_q^+}\sum_{j\in\mathcal{I}_q^-}\ell(\Delta R_{ij}),\qquad\mathcal{L}_g= \frac{\sum_{q\in\mathcal{A}_g}\omega_q\mathcal{L}_q}{\sum_{q\in\mathcal{A}_g}\omega_q},
\label{eq:ordinal_loss}
\end{equation}
where $\mathcal{A}_g$ is the set of rating relations present in group $g$, $\omega_{2\succ0}=1$, and the other relation weights are $0.5$. This two-level reduction makes the contribution of a group depend only on $\omega_q$, invariant to the number of preference pairs and to the group size. Both the policy and frozen reference are initialized from the unified pre-training checkpoint. We train this stage for $104$ optimizer updates and take this as the end point of the stage.

\subsubsection{Generation Reinforcement Learning}
\label{sssec:generation_rl}
Generation reinforcement learning improves two task families with different reward structures. For \textbf{zero-shot TTS}, we optimize content correctness and speaker similarity using off-the-shelf ASR and speaker-verification models. For \textbf{instruction TTS}, we optimize agreement with the requested speaking style using a dedicated style-consistency model. This second stage also mitigates the generation-quality trade-off introduced by editing-oriented preference optimization.

\paragraph{Prompt Construction.}
We construct a frozen pool of $10{,}000$ zero-shot prompts from the training set, with $5{,}000$ Chinese and $5{,}000$ English examples. References are stratified into duration ranges of $1$--$4$\,s, $4$--$8$\,s, and $8$--$15$\,s. Following FlowTTS-GRPO~\cite{wang2026flowttsgrpo}, we augment the pool with hard texts containing local word repetition, sparse multi-word repetition, or whole-sentence repetition, each paired with the original reference waveform. For instruction-following TTS, we categorize the instructions by their controllable dimensions and assess the base model's per-dimension style consistency. Guided by this diagnosis, we deliberately over-sample the dimensions on which the model is weakest, and curate an additional pool of $5{,}000$ instruction-following prompts so that optimization pressure concentrates where the model is least reliable.

\paragraph{Flow-GRPO Optimization.}
Following Flow-GRPO~\cite{liu2025flowgrpo}, we reformulate sampling as an SDE with the same marginal distributions, yielding Gaussian transition kernels along the denoising trajectory. Candidates in a GRPO group are generated from different noise trajectories, providing the exploration required for relative policy optimization. Following MixGRPO~\cite{li2025mixgrpo}, stochastic sampling and gradient computation are restricted to a contiguous six-step window in the low-SNR portion of the first half of the trajectory; the remaining steps use deterministic ODE updates to limit variance and computational cost. The policy objective follows the clipped form of GRPO~\cite{shao2024deepseekmath}, augmented with a KL penalty of weight $\beta_{\mathrm{KL}}$ toward the frozen reference policy.

\paragraph{Reward Design.}
For zero-shot TTS, content correctness is judged by off-the-shelf speech recognition models. Errors caused by Chinese homophones cannot be removed by reinforcement learning, so we distinguish a tolerant error rate $E$ from a strict error rate $E^{\text{s}}$. A homophone substitution does not indicate a pronunciation failure, so the tolerant rate decides whether a candidate counts as fully correct, while the strict rate provides a finer-grained ranking signal among candidates that already pass this check, and is fused with the negative log-likelihood of the target text into a content score $R_i^{\mathrm{content}}$. Let $\gamma\in[0,1]$ denote the interpolation weight between the two terms and $\tau>0$ a temperature controlling the sensitivity of the error term: \begin{equation} R_i^{\mathrm{content}}=(1-\gamma)\Big(1+4e^{-E_i^{\text{s}}/\tau}\Big)+\gamma\Big(1+4e^{-\text{NLL}_i}\Big). \label{eq:content_score} \end{equation}
Speaker similarity is measured as the cosine similarity between the speaker embeddings of the candidate and reference audio, where $\psi(\cdot)$ denotes the speaker encoder:
\begin{equation}
R_i^{\mathrm{spk}}=\cos\big(\psi(y_i),\psi(y_{\text{ref}})\big).
\label{eq:speaker_sim}
\end{equation}

To prevent high speaker similarity from compensating for incorrect content, $R_i^{\mathrm{content}}$ is standardized over the entire candidate group, whereas $R_i^{\mathrm{spk}}$ is standardized only among candidates satisfying $E_i=0$. This constrained construction closes a common reward-hacking path in weighted reward fusion~\cite{wang2026flowttsgrpo,zhang2026swantale}.

For instruction TTS, a dedicated style-consistency model judges whether a candidate matches the requested attributes. We collect 30,000 balanced 1:1 positive/negative samples as supervision, and train a multimodal reward model built based on Qwen2.5-Omni-7B~\cite{Qwen2.5-Omni} to reproduce style-consistency judgement. In practice, we query the judge $V$ times and use majority voting:
\begin{equation}
    R_i^{\mathrm{style}}
    = \mathbb{1}\!\left[
        \sum_{v=1}^{V}
        \mathbb{1}[\operatorname{judge}_v(\mathbf{y}_i)=\mathrm{consistent}]
        > V/2
      \right].
    \label{eq:style_reward}
\end{equation}
The style-task advantage is the standard group-wise z-score used in GRPO~\cite{shao2024deepseekmath}.

\paragraph{Optimization.}
We initialize the policy from the editing-preference checkpoint, sample $G=16$ candidates per prompt, and train for $500$ optimizer updates. Rollouts use $500$ sampling steps, classifier-free guidance of $2.0$, a sway coefficient of $-1.0$, and $\lambda=0.7$. We optimize with AdamW ($\beta=(0.9,0.95)$, weight decay $10^{-4}$), a learning rate of $5\times10^{-5}$, and gradient clipping at $1.0$. The KL coefficient is initialized to $\beta_{\mathrm{KL}}=0.12$ and adjusted proportionally toward a target KL of $\mathcal{D}^{\star}=10^{-3}$ within the interval $[0.08,0.5]$.

\section{Model Acceleration}
\label{sec:model-acceleration}
The full post-trained model requires iterative flow sampling and classifier-free guidance (CFG)\cite{cfg}, making inference expensive across its broad generation and editing capabilities. We therefore distill it into a four-step, CFG-free student through two stages: consistency initialization provides a stable few-step starting point, and task-routed Decoupled DMD improves distribution matching while preserving separation ability. Both stages use data sampled from the same distribution as unified pre-training. The full post-trained model serves as the teacher throughout acceleration.

\paragraph{Consistency Initialization.}
Following the common practice in few-step video diffusion distillation of initializing the student before distribution matching stage~\cite{zhao2026causalforcingpp,zheng2026causalrcm,li2026distillalign,yin2025causvid}, we initialize the student from the teacher and perform trajectory-level consistency distillation\cite{cm} under teacher guidance. The student is trained to map any noisy state $\mathbf{z}_t$ directly to the endpoint $\mathbf{z}_1$ of its sampling trajectory: for neighboring timesteps $t<t'$, the frozen teacher first advances $\mathbf{z}_t$ to $\mathbf{z}_{t'}$ with a single CFG-guided velocity step, and the student's predictions of $\mathbf{z}_1$ from $\mathbf{z}_t$ and from $\mathbf{z}_{t'}$ are matched, with a stop-gradient on the latter so that gradients flow only through the prediction at $t$. This stage adapts the student to few-step flow integration, yielding a stable initialization with preliminary four-step generation and editing capability. We also compared other trajectory-level distillation for initialization, including the ODE regression used in CausVid~\cite{yin2025causvid} and MeanFlow\cite{geng2025meanflow}. At matched update counts, consistency initialization consistently performs best. We therefore keep consistency distillation.

\paragraph{Decoupled DMD with APG.}
The second stage adopts Decoupled DMD~\cite{liu2026decoupleddmd}, which has been validated at scale in Z-Image~\cite{zimage2025}. Decoupled DMD separates the student update into two complementary components. CFG Augmentation (CA) transfers the teacher's conditional guidance to the student, whereas Distribution Matching (DM) aligns the student distribution with the teacher distribution. The two components are evaluated on independently re-noised student predictions so that guidance transfer and distribution matching are not tied to the same noise level.

The consistency checkpoint initializes both the student and the fake-score model, while the frozen post-trained teacher acts as the real-score model. We observe that directly using teacher CFG targets in the CA branch can expose the few-step student to over-saturated predictions, causing overshoot and audible clipping. We therefore replace CFG in this branch with adaptive projected guidance (APG)~\cite{sadat2025apg}, which suppresses excessive guidance components while retaining instruction adherence.

\paragraph{Task-Routed Decoupled DMD.}
We empirically observe that applying Decoupled DMD uniformly across tasks degrades multi-speaker and vocal separation, with some student outputs regressing toward the unprocessed mixture. We attribute this behavior to a distribution mismatch: re-noised erroneous separation outputs can fall outside the teacher's training distribution, causing the teacher field to favor globally plausible but insufficiently separated audio. To preserve separation ability, we route separation examples to a supervised clean-prediction objective and exclude them from both sides of the DMD update.

For sample $i$ with task label $\tau_i$, let $\widehat{\mathbf{z}}_{1,i}$ denote the student's clean-endpoint prediction and $\mathbf{z}_{1,i}$ the ground-truth clean latent. The routed objectives are
\begin{equation}
    \left(\mathcal{L}_{\theta}^{(i)},\mathcal{L}_{\phi}^{(i)}\right)
    =
    \begin{cases}
        \left(
          \left\|\widehat{\mathbf{z}}_{1,i}-\mathbf{z}_{1,i}\right\|_2^2,
          0
        \right),
        & \tau_i\in\mathcal{S},\\[3pt]
        \left(
          \mathcal{L}_{\mathrm{DMD}}^{(i)},
          \mathcal{L}_{\mathrm{fake}}^{(i)}
        \right),
        & \tau_i\notin\mathcal{S},
    \end{cases}
    \label{eq:task-routing-dmd}
\end{equation}
where $\mathcal{L}_{\theta}^{(i)}$ updates the student and $\mathcal{L}_{\phi}^{(i)}$ updates the fake-score model. For routed separation examples, clean-latent regression updates only the student, while the zero in the paired objective excludes the sample from fake-score training. All remaining examples use the Decoupled DMD student loss $\mathcal{L}_{\mathrm{DMD}}^{(i)}$ and the online flow-matching loss $\mathcal{L}_{\mathrm{fake}}^{(i)}$, which keeps the fake-score model aligned with the evolving student distribution.

\paragraph{Training Configuration.} Both stages use the same data distribution and batch size as unified pre-training. We train on $256$ GPUs with fused AdamW, a constant learning rate of $10^{-5}$, and global gradient clipping at $1.0$. Consistency initialization runs for $500$ updates; the student is initialized from the teacher, and teacher targets use CFG with guidance scale $2.0$. The second stage initializes the student and fake-score model from the consistency checkpoint and freezes the teacher as the real-score model. We perform $2{,}500$ student updates and $5$ fake-score updates for every student update. The CA branch uses APG with guidance scale $4.0$ and $\eta=0$. No additional adversarial discriminator or adversarial loss is introduced.

\section{Model Inference}
\label{sec:model-inference}
The inference pipeline contains two stages. For free-form user requests, a Prompt Enhancer (PE) identifies the task, rewrites the request into a model-oriented instruction, prepares optional input audio, and estimates the output duration. The resulting instruction and optional audio context are then passed to \modelname for conditional flow sampling. Canonical instructions that already follow the supported format can bypass PE and be provided directly to the model.

\subsection{Prompt Enhancer}
Free-form requests can vary substantially in wording, omit required parameters, or refer implicitly to the input audio. PE provides a task-aware interface that converts these requests into explicit instructions closer to the training distribution while preserving all user-specified content.

\paragraph{Task Routing and Instruction Rewriting.}
Given a request $q$ and optional input audio $a$, PE obtains an ASR transcript and language prediction when audio is available. The request, audio, and ASR context are then processed by a capable language model~\cite{tencenthyteam2026hy3, deepseekai2026deepseekv4, openai2026gpt5.6} to identify the target task and extract its arguments and control parameters. Audio-derived context helps resolve implicit references, distinguish operations with similar wording, and support duration estimation.

The extracted parameters are validated against the ranges represented during training. Colloquial or continuous descriptions of speaking rate, loudness, and pitch are mapped to supported discrete values, while invalid or unsupported requests are rejected before acoustic inference. PE then renders the validated request using a task-specific template or a curated instruction formulation. User-provided synthesis text and replacement content are preserved, while task descriptions may be lightly rewritten or expanded. Text normalization is applied when required for speech generation.

\paragraph{Audio Preparation and Duration Estimation.}
For audio-conditioned tasks, PE applies lightweight, task-dependent preprocessing, including leading- and trailing-silence handling, resampling, and simple level normalization when the input distribution differs from that used during training. These operations standardize the model input without altering content that should be preserved. PE also estimates the target duration before flow sampling. Let $T_{\mathrm{in}}$ be the input-audio duration, let $B(x)$ denote the UTF-8 byte length of text $x$, following the duration heuristic used in F5-TTS~\cite{f5tts}. The task-dependent estimate is
\begin{equation}
T_{\mathrm{out}} =
\begin{cases}
T_{\mathrm{in}}
\dfrac{B(x_{\mathrm{target}})}
      {B(x_{\mathrm{source}})},
& \text{zero-shot TTS and content editing}, \\[7pt]
\dfrac{T_{\mathrm{in}}}{s},
& \text{speed editing with multiplier } s, \\[7pt]
\kappa_e T_{\mathrm{in}},
& \text{emotion editing}, \\[4pt]
T_{\mathrm{in}}+\Delta_{\mathrm{nv}},
& \text{nonverbal editing}, \\[4pt]
f_{\mathrm{LLM}}\!\left(
\widehat{T}_{\mathrm{base}},
d_{\mathrm{style}},
x_{\mathrm{target}}
\right),
& \text{instruct TTS}, \\[4pt]
T_{\mathrm{in}},
& \text{duration-preserving operations}.
\end{cases}
\label{eq:pe-duration}
\end{equation}

The emotion factor $\kappa_e$ and nonverbal offset $\Delta_{\mathrm{nv}}$ are estimated from the corresponding training data. For instruct TTS, the base estimate $\widehat{T}_{\mathrm{base}}=\rho_{\ell}B(x_{\mathrm{target}})$ uses a language-dependent byte-rate coefficient $\rho_{\ell}$; the language model then adjusts this estimate using the requested style and output text. For zero-shot TTS, $x_{\mathrm{source}}$ is obtained internally by transcribing the reference audio, so the user is not required to provide a reference transcript. For speech and lyric content editing, the duration is scaled using the UTF-8 byte-length ratio between the target and source content.

\subsection{Inference Strategy and Configuration}
When input audio is provided, it is converted to mono, resampled to $24$\,kHz, and encoded by the VAE as an acoustic reference. In parallel, the MLLM jointly processes the enhanced instruction and input audio to construct the semantic condition. For text-only generation, the semantic condition is derived from the instruction alone and no acoustic reference is used. The estimated duration determines the target latent length at the VAE rate of $50$\,Hz. Sampling starts from Gaussian noise of this length, and the generated latent is decoded by the VAE into the output waveform.

All evaluation-time settings for \modelname and \modelname-Flash are consolidated in \cref{tab:inference-configuration}. Both variants use the same prompt enhancement, audio preparation, duration estimation, latent representation, and VAE decoder. The full model uses its EMA checkpoint with classifier-free guidance, whereas the Flash variant uses the task-routed distilled checkpoint and requires no guidance.

\begin{table}[ht]
\centering
\small
\setlength{\tabcolsep}{5pt}
\renewcommand{\arraystretch}{1.08}
\caption{Inference configurations for \modelname and \modelname-Flash.}
\label{tab:inference-configuration}
\begin{tabular}{lcc}
\toprule
\textbf{Configuration} & \textbf{\modelname} & \textbf{\modelname-Flash} \\
\midrule
Checkpoint & EMA post-trained & Task-routed distilled \\
Numerical precision & \multicolumn{2}{c}{bfloat16} \\
ODE solver & \multicolumn{2}{c}{Euler} \\
Function evaluations & $32$ & $4$ \\
CFG scale & $2.0$ & None \\
Sway coefficient & \multicolumn{2}{c}{$-1.0$} \\
Output sample rate & \multicolumn{2}{c}{$24$\,kHz} \\
VAE latent space & \multicolumn{2}{c}{$64$ dim, $50$\,Hz} \\
Wall-clock speedup & $1.0\times$ & $4.5\times$ \\
\bottomrule
\end{tabular}
\end{table}

The reported $4.5\times$ speedup compares the four-step, CFG-free \modelname-Flash sampler against the $32$-NFE, CFG-$2.0$ \modelname sampler under the same hardware, output duration, and batch size. Unless otherwise stated, these configurations are used for the corresponding benchmark results.

\section{Performance}

We evaluate \modelname and \modelname-Flash from three complementary perspectives: reconstruction fidelity, speech generation, and speech editing.
\Cref{tab:sota-summary} provides a compact comparison with representative prior systems, while \Cref{app:detailed-evaluation-results} reports additional metrics, operation-level results, and broader baseline sets where available.
Across these evaluations, both variants achieve consistently strong performance in zero-shot and instruction-controlled generation, general speech editing, and signal-level speech processing.
The full model generally provides stronger linguistic accuracy and edit fidelity, whereas the Flash variant retains competitive instruction following and often offers better perceptual quality under accelerated inference.
We analyze VAE reconstruction, generation, and editing in \Cref{sec:vae-reconstruction,sec:generation-ability,sec:editing-ability}, respectively.

\begin{table*}[t]
\centering
\caption{Summary of selected speech generation, editing, enhancement, and separation benchmarks. Each block retains its native metrics. The best result among the displayed systems in each column is shown in bold.}
\vspace{2mm}
\setlength{\tabcolsep}{3pt}
\renewcommand{\arraystretch}{1.06}
\newcommand{\sotabest}[1]{\textbf{#1}}
\newcommand{\sotaauk}{\textcolor[HTML]{4075FF}{\modelname}}
\newcommand{\sotaaukflash}{\textcolor[HTML]{06BAFF}{\modelname-Flash}}
\newcommand{\sotaaukresult}[1]{\cellcolor[HTML]{E6EEFF}#1}
\newcommand{\sotaaukflashresult}[1]{\cellcolor[HTML]{E6F8FF}#1}
\newcommand{\sotavalue}[1]{\makebox[2.4em][r]{#1}}
\newcommand{\sotasep}{\,\textnormal{\textbar{}}\,}
\newcommand{\sotatwo}[2]{\sotavalue{#1}\sotasep\sotavalue{#2}}
\newcommand{\sotathree}[3]{\sotavalue{#1}\sotasep\sotavalue{#2}\sotasep\sotavalue{#3}}
\newcommand{\sotafour}[4]{\sotavalue{#1}\sotasep\sotavalue{#2}\sotasep\sotavalue{#3}\sotasep\sotavalue{#4}}
\newcommand{\sotafive}[5]{\sotavalue{#1}\sotasep\sotavalue{#2}\sotasep\sotavalue{#3}\sotasep\sotavalue{#4}\sotasep\sotavalue{#5}}
\newcommand{\sotasix}[6]{\sotavalue{#1}\sotasep\sotavalue{#2}\sotasep\sotavalue{#3}\sotasep\sotavalue{#4}\sotasep\sotavalue{#5}\sotasep\sotavalue{#6}}
\resizebox{\textwidth}{!}{
\begin{tabular}{@{}llccc@{}}
\toprule
\textbf{Task} & \textbf{Dataset} & \textbf{Prior SOTA} & \textbf{Metrics} & \textbf{Results} \\
\midrule
\multirow{15}{*}{\shortstack[c]{\textbf{Generation} \\ \textbf{Ability}}}
& \multirow{10}{*}{\shortstack[l]{\textbf{Seed-TTS-Eval} \\ \textit{en\sotasep zh\sotasep zh-hard\sotasep avg}}}
& Qwen3-TTS & \multirow{5}{*}{WER$\downarrow$} & \sotafour{1.23}{1.22}{6.76}{3.07} \\
& & Seed-TTS & & \sotafour{2.25}{1.12}{7.59}{3.65} \\
& & VoxCPM2 & & \sotafour{1.84}{\sotabest{0.97}}{8.13}{3.65} \\
& & \sotaaukflash & & \sotaaukflashresult{\sotafour{1.03}{1.10}{6.43}{2.85}} \\
& & \sotaauk & & \sotaaukresult{\sotafour{\sotabest{1.02}}{1.02}{\sotabest{5.91}}{\sotabest{2.65}}} \\
\cmidrule(lr){3-5}
& & Qwen3-TTS & \multirow{5}{*}{SIM$\uparrow$} & \sotafour{0.717}{0.770}{0.748}{0.745} \\
& & Seed-TTS & & \sotafour{0.762}{0.796}{0.776}{0.778} \\
& & VoxCPM2 & & \sotafour{0.753}{0.795}{0.753}{0.767} \\
& & \sotaaukflash & & \sotaaukflashresult{\sotafour{0.772}{\sotabest{0.814}}{\sotabest{0.784}}{0.790}} \\
& & \sotaauk & & \sotaaukresult{\sotafour{\sotabest{0.788}}{\sotabest{0.814}}{0.782}{\sotabest{0.795}}} \\
\cmidrule(lr){2-5}
& \multirow{5}{*}{\shortstack[l]{\textbf{InstructTTSEval} \\ \textit{DSD-ZH\sotasep DSD-EN}}}
& Qwen3-TTS-VD & \multirow{5}{*}{ACC$\uparrow$} & \sotatwo{81.10}{\sotabest{82.40}} \\
& & Mimo-Audio & & \sotatwo{74.30}{77.60} \\
& & MOSS-VoiceGenerator & & \sotatwo{80.00}{82.00} \\
& & \sotaaukflash & & \sotaaukflashresult{\sotatwo{78.80}{\sotabest{82.40}}} \\
& & \sotaauk & & \sotaaukresult{\sotatwo{\sotabest{83.37}}{81.60}} \\
\midrule
\multirow{26}{*}{\shortstack[c]{\textbf{General Speech} \\ \textbf{Editing}}}
& \multirow{4}{*}{\textbf{MMAE(Speech)}}
& Step-Audio-EditX & \multirow{4}{*}{IFR$\uparrow$\sotasep CR$\uparrow$\sotasep EMR$\uparrow$} & \sotathree{43.52}{77.27}{4.69} \\
& & Ming-UniAudio & & \sotathree{34.13}{76.01}{7.04} \\
& & \sotaaukflash & & \sotaaukflashresult{\sotathree{46.62}{86.41}{\sotabest{13.85}}} \\
& & \sotaauk & & \sotaaukresult{\sotathree{\sotabest{48.23}}{\sotabest{88.11}}{12.44}} \\
\cmidrule(lr){2-5}
& \multirow{4}{*}{\shortstack[l]{\textbf{SpeechEditBench} \\ \textit{Content\sotasep Emotion\sotasep Prosody\sotasep} \\ \textit{Paralinguistic\sotasep Acoustic}}}
& Ming-UniAudio & \multirow{4}{*}{ACC$\uparrow$} & \sotafive{76.46}{3.43}{26.50}{11.25}{25.85} \\
& & Step-Audio-EditX & & \sotafive{16.50}{7.71}{20.13}{31.25}{22.89} \\
& & \sotaaukflash & & \sotaaukflashresult{\sotafive{87.50}{6.29}{70.00}{\sotabest{39.25}}{30.26}} \\
& & \sotaauk & & \sotaaukresult{\sotafive{\sotabest{91.83}}{\sotabest{9.94}}{\sotabest{71.33}}{38.50}{\sotabest{37.07}}} \\
\cmidrule(lr){2-5}
& \multirow{12}{*}{\shortstack[l]{\textbf{Ming-Freeform-Audio-Edit} \\ \textit{Semantic Editing} \\ \textit{Basic-ZH\sotasep Full-ZH} \\ \textit{Basic-EN\sotasep Full-EN}}}
& Ming-UniAudio & \multirow{3}{*}{WER$\downarrow$} & \sotafour{6.61}{10.46}{10.16}{14.28} \\
& & \sotaaukflash & & \sotaaukflashresult{\sotafour{4.87}{3.34}{4.71}{4.84}} \\
& & \sotaauk & & \sotaaukresult{\sotafour{\sotabest{4.62}}{\sotabest{3.09}}{\sotabest{3.85}}{\sotabest{3.96}}} \\
\cmidrule(lr){3-5}
& & Ming-UniAudio & \multirow{3}{*}{ACC$\uparrow$} & \sotafour{86.21}{79.62}{71.14}{70.98} \\
& & \sotaaukflash & & \sotaaukflashresult{\sotafour{\sotabest{92.01}}{91.21}{84.00}{82.98}} \\
& & \sotaauk & & \sotaaukresult{\sotafour{91.95}{\sotabest{91.47}}{\sotabest{85.47}}{\sotabest{85.25}}} \\
\cmidrule(lr){3-5}
& & Ming-UniAudio & \multirow{3}{*}{SIM$\uparrow$} & \sotafour{0.81}{0.82}{0.78}{0.77} \\
& & \sotaaukflash & & \sotaaukflashresult{\sotafour{\sotabest{0.88}}{\sotabest{0.89}}{\sotabest{0.89}}{\sotabest{0.89}}} \\
& & \sotaauk & & \sotaaukresult{\sotafour{\sotabest{0.88}}{0.88}{0.88}{0.88}} \\
\cmidrule(lr){3-5}
& & Ming-UniAudio & \multirow{3}{*}{no-edit WER$\downarrow$} & \sotafour{6.55}{8.78}{20.41}{24.15} \\
& & \sotaaukflash & & \sotaaukflashresult{\sotafour{5.01}{3.32}{16.55}{16.76}} \\
& & \sotaauk & & \sotaaukresult{\sotafour{\sotabest{4.77}}{\sotabest{3.11}}{\sotabest{15.84}}{\sotabest{15.99}}} \\
\cmidrule(lr){2-5}
& \multirow{6}{*}{\shortstack[l]{\textbf{Ming-Freeform-Audio-Edit} \\ \textit{Acoustic Editing} \\ \textit{ZH\sotasep EN}}}
& Ming-UniAudio & \multirow{3}{*}{WER$\downarrow$} & \sotatwo{5.01}{10.75} \\
& & \sotaaukflash & & \sotaaukflashresult{\sotatwo{2.40}{4.45}} \\
& & \sotaauk & & \sotaaukresult{\sotatwo{\sotabest{2.02}}{\sotabest{3.48}}} \\
\cmidrule(lr){3-5}
& & Ming-UniAudio & \multirow{3}{*}{SIM$\uparrow$} & \sotatwo{0.63}{0.54} \\
& & \sotaaukflash & & \sotaaukflashresult{\sotatwo{\sotabest{0.79}}{\sotabest{0.75}}} \\
& & \sotaauk & & \sotaaukresult{\sotatwo{0.78}{0.74}} \\
\midrule
\multirow{19}{*}{\shortstack[c]{\textbf{Speech Enhancement} \\ \textbf{and Separation}}}
& \multirow{5}{*}{\textbf{DNS Challenge}}
& SAM-Audio-Large & \multirow{5}{*}{DNSMOS-OVRL$\uparrow$\sotasep UTMOS$\uparrow$\sotasep dWER$\downarrow$\sotasep SIM$\uparrow$} & \sotafour{3.29}{3.78}{9.67}{0.96} \\
& & AnyEnhance & & \sotafour{\sotabest{3.41}}{3.76}{7.71}{0.98} \\
& & RE-USE & & \sotafour{3.38}{3.69}{3.31}{0.98} \\
& & \sotaaukflash & & \sotaaukflashresult{\sotafour{3.38}{\sotabest{4.05}}{2.93}{\sotabest{0.99}}} \\
& & \sotaauk & & \sotaaukresult{\sotafour{3.35}{3.86}{\sotabest{2.66}}{\sotabest{0.99}}} \\
\cmidrule(lr){2-5}
& \multirow{5}{*}{\textbf{CHiME-4}}
& SAM-Audio-Large & \multirow{5}{*}{DNSMOS-OVRL$\uparrow$\sotasep UTMOS$\uparrow$\sotasep WER$\downarrow$} & \sotathree{3.14}{3.38}{17.30} \\
& & AnyEnhance & & \sotathree{3.19}{3.24}{27.41} \\
& & RE-USE & & \sotathree{3.33}{3.44}{10.71} \\
& & \sotaaukflash & & \sotaaukflashresult{\sotathree{\sotabest{3.35}}{\sotabest{3.91}}{\sotabest{7.84}}} \\
& & \sotaauk & & \sotaaukresult{\sotathree{3.28}{3.72}{7.98}} \\
\cmidrule(lr){2-5}
& \multirow{5}{*}{\textbf{Libri2Mix}}
& MossFormer2-SS & \multirow{5}{*}{DNSMOS-OVRL$\uparrow$\sotasep UTMOS$\uparrow$\sotasep WER$\downarrow$\sotasep SIM$\uparrow$} & \sotafour{3.24}{3.66}{9.34}{\sotabest{0.96}} \\
& & SAM-Audio-Large & & \sotafour{2.87}{2.84}{57.22}{0.86} \\
& & Sidon (Dialogue) & & \sotafour{3.08}{2.61}{51.79}{0.87} \\
& & \sotaaukflash & & \sotaaukflashresult{\sotafour{\sotabest{3.32}}{\sotabest{4.03}}{10.07}{\sotabest{0.96}}} \\
& & \sotaauk & & \sotaaukresult{\sotafour{3.28}{3.87}{\sotabest{9.12}}{\sotabest{0.96}}} \\
\cmidrule(lr){2-5}
& \multirow{4}{*}{\textbf{VCTKSR}}
& AudioSR & \multirow{4}{*}{DNSMOS-OVRL$\uparrow$\sotasep UTMOS$\uparrow$\sotasep WER$\downarrow$\sotasep SIM$\uparrow$} & \sotafour{3.08}{3.13}{4.72}{0.92} \\
& & Resemble-Enhance & & \sotafour{3.18}{3.59}{15.94}{0.95} \\
& & \sotaaukflash & & \sotaaukflashresult{\sotafour{\sotabest{3.25}}{\sotabest{4.05}}{\sotabest{2.92}}{0.96}} \\
& & \sotaauk & & \sotaaukresult{\sotafour{3.21}{3.93}{3.06}{\sotabest{0.97}}} \\
\bottomrule
\end{tabular}%
}
\vspace{-2mm}
\label{tab:sota-summary}\end{table*}

\subsection{VAE Reconstruction Results}
\label{sec:vae-reconstruction}
We evaluate the reconstruction quality of \modelname-VAE across three representative audio domains: speech, general audio, and music. Specifically, the evaluation uses Seed-TTS-Eval\cite{seedtts} for speech, 2,000 randomly sampled clips from AudioSet\cite{audioset} for general audio, and the MUSDB18-HQ\cite{MUSDB18HQ} test set for music. The compared models include MiniMax-H3-AudioVAE\cite{minimax_h3_audio_vae}, Ming-Omni-TTS\cite{mingomnitts}, Stable-Audio-3-SAME-L\cite{stableaudio3-same}, and MMAudio-VAE\cite{mmaudio}. Reconstruction fidelity is measured using PESQ, STOI, mel-spectrogram distance (Mel Dist), and multi-resolution STFT distance (STFT).

As shown in \cref{tab:vae-reconstruction}, \modelname-VAE achieves the best result
on all four metrics across all three domains. These results demonstrate that
\modelname-VAE consistently preserves perceptual quality, intelligibility, and
spectral detail across speech, general audio, and music.

\begin{table*}[t]
\centering
\scriptsize
\setlength{\tabcolsep}{4pt}
\renewcommand{\arraystretch}{1.05}
\caption{Comparison of VAE reconstruction quality across speech, general
audio, and music. The best and second-best results within each domain are shown in
bold and underlined, respectively.}
\label{tab:vae-reconstruction}
\vspace{2mm}
\begin{tabular}{@{}lllcccc@{}}
\toprule
\textbf{Domain} & \textbf{Test Set} & \textbf{Method} &
\textbf{PESQ}$\uparrow$ & \textbf{STOI}$\uparrow$ &
\textbf{Mel Dist}$\downarrow$ &
\textbf{STFT}$\downarrow$ \\
\midrule
\multirow{5}{*}{Speech} & \multirow{5}{*}{Seed-TTS-Eval}
& MiniMax-H3-AudioVAE & \underline{3.633} & \underline{0.968} &
0.695 & \underline{1.615} \\
& & Ming-omni-tts & 2.583 & 0.926 & 1.014 & 1.923 \\
& & Stable-Audio-3-SAME-L & 2.969 & 0.951 & 0.954 & 1.819 \\
& & MMAudio-VAE & 2.696 & 0.937 & \underline{0.619} & 1.627 \\
& & \textbf{\modelname-VAE} & \textbf{4.143} & \textbf{0.982} &
\textbf{0.574} & \textbf{1.457} \\
\midrule
\multirow{5}{*}{Audio} & \multirow{5}{*}{AudioSet}
& MiniMax-H3-AudioVAE & 2.835 & \underline{0.792} &
0.766 & \underline{2.119} \\
& & Ming-omni-tts & 1.817 & 0.650 & 1.121 & 3.389 \\
& & Stable-Audio-3-SAME-L & \underline{2.901} & 0.779 & 1.052 & 3.264 \\
& & MMAudio-VAE & 2.455 & 0.770 & \underline{0.635} & 2.216 \\
& & \textbf{\modelname-VAE} & \textbf{3.833} & \textbf{0.900} &
\textbf{0.571} & \textbf{1.842} \\
\midrule
\multirow{5}{*}{Music} & \multirow{5}{*}{MUSDB18-HQ}
& MiniMax-H3-AudioVAE & 2.801 & 0.847 & 0.697 & 1.848 \\
& & Ming-omni-tts & 1.634 & 0.708 & 0.941 & 2.315 \\
& & Stable-Audio-3-SAME-L & \underline{3.051} & \underline{0.876} &
0.701 & \underline{1.795} \\
& & MMAudio-VAE & 1.753 & 0.775 & \underline{0.661} & 1.826 \\
& & \textbf{\modelname-VAE} & \textbf{3.884} & \textbf{0.927} &
\textbf{0.525} & \textbf{1.721} \\
\bottomrule
\end{tabular}
\end{table*}

\subsection{Generation Ability}
\label{sec:generation-ability}
We evaluate two complementary generation capabilities: zero-shot voice cloning on Seed-TTS-Eval~\cite{seedtts} and instruction-controlled synthesis on InstructTTSEval~\cite{instructttseval}. Seed-TTS-Eval measures linguistic accuracy and speaker preservation across English, Chinese, and Chinese hard-text subsets, while InstructTTSEval evaluates control over acoustic parameters, descriptive styles, and role-playing instructions in both languages. For both benchmarks, inference uses only benchmark-specific instruction templates and output-duration estimation, without additional prompt enhancement. We run each evaluation three times and report the mean score.

\paragraph{Zero-Shot TTS.}

\modelname achieves the lowest average recognition error and the highest average speaker similarity among the compared systems, with an average error of 2.65\% and a SIM of 0.795.
The comparison in \Cref{tab:sota-summary} includes Qwen3-TTS~\cite{qwen3tts}, Seed-TTS~\cite{seedtts}, and VoxCPM2~\cite{voxcpm2}, while \Cref{tab:seed-tts-eval} provides a broader baseline set.
Compared with Qwen3-TTS, the strongest baseline in average recognition error, \modelname reduces the error from 3.07\% to 2.65\%.
It also improves the average SIM over Seed-TTS, the strongest baseline on this metric, from 0.778 to 0.795.
\modelname-Flash remains competitive, achieving an average recognition error of 2.85\% and a SIM of 0.790.
On test-en, \modelname achieves the lowest WER of 1.02\%; on test-zh-hard, it obtains the lowest recognition error of 5.91\%, while \modelname-Flash achieves the highest SIM of 0.784.

\paragraph{Instruct TTS.}
On the DSD split, \modelname achieves the best Chinese accuracy of 83.37\%, outperforming Qwen3-TTS-VD~\cite{qwen3tts} by 2.27 percentage points.
\modelname-Flash achieves 82.40\% on English DSD, tying Qwen3-TTS-VD for the best result among the systems summarized in \Cref{tab:sota-summary}.
The complete results in \Cref{tab:instruct-tts-eval}, which also include Mimo-Audio~\cite{mimoaudio} and MOSS-VoiceGenerator~\cite{mossvoicegenerator}, show that the full model outperforms the Flash variant on all three Chinese metrics, whereas the Flash variant's main advantage is its stronger English DSD result.

\subsection{Editing Ability}
\label{sec:editing-ability}
We evaluate editing-related capabilities in two complementary regimes.
The first tests whether a model can execute natural-language editing instructions while preserving attributes outside the requested change.
The second regime evaluates signal-level processing using DNS Challenge 2020~\cite{dnschallenge2020} and CHiME-4~\cite{chime4} for enhancement, Libri2Mix~\cite{libri2mix} for two-speaker separation, and VCTK-SR~\cite{vctksr} for speech super-resolution.

\subsubsection{General Speech Editing}

We evaluate general speech editing on three complementary benchmarks.
For MMAE~\cite{ma2026mmae}, we use its speech subset to assess rubric-based instruction following and preservation.
SpeechEditBench~\cite{zhang2026speecheditbench} evaluates joint success across five editing categories, while Ming-Freeform-Audio-Edit~\cite{yan2025ming} provides operation-level evaluations of semantic and acoustic editing.
\paragraph{MMAE-Speech.}
IFR measures the average success rate on instruction-following rubrics, CR measures consistency on attributes unrelated to the requested edit, and EMR is the percentage of samples that satisfy all instruction-following and consistency rubrics.
\modelname achieves the highest IFR and CR scores, reaching 48.23\% and 88.11\%, respectively, while \modelname-Flash achieves the highest EMR of 13.85\%.
Relative to Step-Audio-EditX~\cite{yan2025step}, the strongest prior baseline on IFR and CR, \modelname improves the two metrics by 4.71 and 10.84 percentage points, respectively.
Relative to Ming-UniAudio~\cite{yan2025ming}, \modelname-Flash improves EMR from 7.04\% to 13.85\%.
These results show that the full model performs better on average instruction following and preservation, while the Flash variant more frequently satisfies all evaluation rubrics simultaneously.

\paragraph{SpeechEditBench.}
Among the dedicated editing models included in \Cref{tab:sota-summary}, \modelname achieves the best results in content, emotion, prosody, and acoustic editing, while \modelname-Flash performs best in paralinguistic editing.
Relative to Ming-UniAudio, \modelname improves the joint success rate from 76.46\% to 91.83\% for content editing and from 26.50\% to 71.33\% for prosody editing.
\modelname-Flash achieves a paralinguistic editing score of 39.25\%, compared with 31.25\% for Step-Audio-EditX.
\paragraph{Ming-Freeform-Audio-Edit.}
The benchmark contains complementary semantic and acoustic editing tracks.
For semantic editing, the summary results average deletion, insertion, and substitution for each Basic/Full and Chinese/English setting.
\modelname achieves the lowest average WER and no-edit WER across all four settings.
Under the more challenging Full setting, compared with Ming-UniAudio, \modelname reduces the average WER from 10.46\% to 3.09\% in Chinese and from 14.28\% to 3.96\% in English.
It also improves editing accuracy from 79.62\% to 91.47\% in Chinese and from 70.98\% to 85.25\% in English.
The operation-level results in \Cref{tab:ming-freeform-semantic-basic,tab:ming-freeform-semantic-full} show that these gains extend across deletion, insertion, and substitution.

For acoustic editing, the summary results average speed, pitch, and volume alteration.
Compared with Ming-UniAudio, \modelname reduces the average WER from 5.01\% to 2.02\% in Chinese and from 10.75\% to 3.48\% in English.
In contrast, \modelname-Flash achieves the highest average SIM scores of 0.79 and 0.75 in Chinese and English, respectively.
The operation-level results in \Cref{tab:ming-freeform-acoustic} show that \modelname achieves the lowest WER for Chinese pitch and volume alteration and for English speed and pitch alteration.
\modelname-Flash achieves the highest or tied-highest SIM in all six language--operation settings.
Overall, the full model provides lower average recognition error, whereas the Flash variant better preserves speaker identity.

\subsubsection{Speech Enhancement and Separation}

The evaluation covers representative task-specific systems, including Resemble-Enhance~\cite{resemble-enhance}, DaSheng~\cite{dasheng}, MossFormer2-SS~\cite{mossformer2}, and AudioSR~\cite{audiosr}, as well as unified or general-purpose systems such as Ming-UniAudio~\cite{yan2025ming}, SAM-Audio-Large~\cite{samaudio}, AnyEnhance~\cite{zhang2025anyenhance}, RE-USE~\cite{fu2026rethinking}, and Sidon~\cite{dialoguesidon}.
The main results are summarized in \Cref{tab:sota-summary}, with complete metrics reported in \Cref{tab:dnschallenge-complete,tab:chime4-complete,tab:libri2mix-complete,tab:vctk-sr-complete}.
Across these tasks, DNSMOS and UTMOS assess perceptual quality, dWER, WER, and PER measure linguistic preservation, and SIM measures speaker-identity preservation.

\paragraph{Speech Enhancement.}
On DNS Challenge, \modelname achieves the lowest dWER of 2.66\%, improving over RE-USE by 0.65 percentage points, while both variants achieve the highest SIM of 0.99.
\modelname-Flash obtains the highest UTMOS score of 4.05 while matching the full model in speaker similarity.
On CHiME-4, \modelname-Flash achieves the lowest WER of 7.84\% and the highest UTMOS score of 3.91; its OVRL score of 3.35 also ties the best result in the complete comparison.
\modelname remains close in recognition accuracy, with a WER of 7.98\%.
Across the two enhancement benchmarks, the Flash variant consistently achieves higher UTMOS, while the best recognition result depends on the dataset.

\paragraph{Speech Separation.}
On Libri2Mix, both variants achieve a SIM of 0.96, tying MossFormer2-SS for the best speaker-similarity result.
\modelname achieves the lowest WER and PER, at 9.12\% and 6.63\%, respectively.
In contrast, \modelname-Flash achieves the highest OVRL and UTMOS scores, at 3.32 and 4.03.
These results reveal a clear trade-off: the full model better preserves linguistic content, whereas the Flash variant achieves higher predicted perceptual quality.

\paragraph{Speech Super-Resolution.}
For speech super-resolution, we apply eight degradation settings to 500 utterances from five held-out VCTK speakers, producing 4,000 test samples. Five subsets evaluate bandwidth extension at effective cutoffs from 2 to 11~kHz, where the 2-kHz condition lies outside the training range and serves as an extrapolation test. The remaining subsets simulate telephone, megaphone, and underwater channels. All inputs and references remain 24-kHz mono waveforms, and we report results averaged uniformly over the eight subsets.

On VCTK-SR, \modelname-Flash achieves the best results on six of the seven reported metrics: OVRL, SIG, BAK, UTMOS, WER, and PER.
It obtains OVRL and UTMOS scores of 3.25 and 4.05, respectively, together with a WER of 2.92\% and a PER of 4.19\%.
Meanwhile, \modelname achieves the highest SIM of 0.97.
These results show that the Flash variant provides the strongest perceptual-quality and recognition results, while the full model provides the strongest speaker preservation.

\subsection{Discovery}

Beyond the benchmark results, the development of \modelname reveals several observations about data construction, capability transfer, and native instruction following.

\paragraph{Cross-Utterance In-Context Learning.}
Our zero-shot TTS data construction differs from the conventional within-utterance setting, where a single recording is divided into prompt and target segments. Instead, we identify distinct utterances from the same speaker and use one utterance as the acoustic prompt and another as the synthesis target. Because the two utterances contain different linguistic content and may exhibit natural variation in prosody and recording conditions, this construction encourages the model to separate speaker-invariant characteristics from utterance-specific factors. In our experiments, cross-utterance training improves both expressiveness and speaker similarity. It also enables a transcript-free interface: the model is conditioned only on the prompt waveform and target text, without requiring the transcript corresponding to the prompt audio.

\paragraph{Emergent Cross-Task and Cross-Lingual Transfer.}
We qualitatively observe capabilities that are not explicitly represented by matched training tasks. First, our whisper data contains only normal-to-whisper or whisper-to-normal editing pairs. Nevertheless, the jointly trained model can also synthesize whispered speech directly from text and a style instruction, suggesting that a transformation learned through editing can transfer to generation. Second, our de-accenting supervision covers Chinese dialects and regional accents only, yet the model can reduce accents in English speech, including English spoken with Indian or Japanese accents, while largely preserving speaker identity. These observations are not substitutes for comprehensive benchmark evaluation, but they suggest that unified training can factorize certain acoustic transformations from the language and task through which they are supervised.

\paragraph{Limits of Native Free-Form Instruction Following.}
We also explored an agent-based data pipeline modified from Audio-Oscar~\cite{duan2026audio} for free-form audio-editing SFT pairs. Although this pipeline increased the linguistic diversity of editing instructions, it did not produce sufficiently robust generalization to arbitrary user requests. In practice, reaching the model's capability ceiling still requires the Prompt Enhancer to identify the intended task, normalize control parameters, and rewrite the request toward the training distribution. A similar dependence on prompt rewriting is commonly observed in image, video, and audio-visual generation systems, indicating a broader gap between possessing a capability and invoking it reliably through unconstrained language. Improving native instruction grounding and compositional generalization, while reducing reliance on explicit task routing and prompt enhancement, remains an important direction for future work.

\section{Conclusion}
We presented \modelname, an open-source foundational model that unifies speech generation and editing through a common instruction-conditioned waveform generation interface. Its training corpus spans five task families and contains approximately $3.03$ billion instruction--audio instances and $1.95$ million hours of effective supervision. A multimodal language model provides semantic conditioning, an audio VAE jointly trained on speech, general audio, and music provides a shared acoustic latent space, and a flex-style Transformer for latent diffusion. Generation-only warm-up, joint generation--editing pre-training, human-feedback preference optimization for open-ended editing, and reward-based reinforcement learning for generation together align the model across this heterogeneous capability set. Task-routed distillation further produces \modelname-Flash, which retains broad capability with four-step inference, no classifier-free guidance, and a $4.5\times$ wall-clock speedup.

Experiments show leading performance on zero-shot and vocie-design speech generation and general instruction-guided editing, together with competitive results on speech restoration tasks. Our qualitative observations also suggest that unified training enables useful transfer across utterances, tasks, and languages. At the same time, robust native understanding of unconstrained editing requests remains incomplete, and the system still benefits from explicit task routing and prompt enhancement. Future work should improve native instruction grounding, compositional generalization, and scalable alignment for open-ended audio transformations while further reducing inference cost. We release the source code and model weights to facilitate future development and research.

\clearpage

\definecolor{leadcolor}{HTML}{4285F4}
\definecolor{corecolor}{HTML}{EA4335}
\definecolor{contribcolor}{HTML}{34A853}
\definecolor{sponsorcolor}{HTML}{FBBC05}
\definecolor{boxbg}{HTML}{F8F9FA}
\definecolor{boxframe}{HTML}{DADCE0}

\section*{Contribution}
\let\thefootnote\relax
\addtocounter{footnote}{-1}  
\addcontentsline{toc}{section}{Contribution}


\vspace{0.4cm}

\noindent{\faStar\;\textcolor{corecolor}{\textbf{Core Contributors}}} \\[4pt]
Ziyang Ma, 
Zhikang Niu, 
Wenming Tu, 
Tianrui Wang,
Ruiqi Yan,
Junxi Liu,
Yanru Huo

\vspace{0.4cm}

\noindent{\faUsers\;\textcolor{contribcolor}{\textbf{Contributors}}}

\vspace{0.2cm}

\def\thefootnote{\arabic{footnote}}\setcounter{footnote}{0}\hspace{0.6em}{\faCogs\;\textbf{Engineering \& Training}\footnote{Nickk Huang, Yang Liu, and Qicong Xie provided support for the VAE, Zeyu Xie and Hui Wang provided the support for sound generation and editing post-training, Haitao Li provided support for InstructTTS post-training, and Zixuan Jiang provided support for the prompt enhancer.}} \\[3pt]
Nickk Huang,
Yang Liu,
Qicong Xie,
Zeyu Xie,
Hui Wang,
Haitao Li,
Zixuan Jiang

\vspace{0.25cm}

\hspace{0.6em}{\faPaintBrush\;\textbf{Data \& Infra}\footnote{Yalin Li, Jie Fang, Yifan Duan, and Zeyue Tian provided support for audio editing data, Guangzheng Li, Haina Zhu, Shuyi Wang, and Jinwen Wang provided support for music editing data, Mingyu Cui provided support for dialect speech data, Tian Tan, Auden, and Sen Liang provided support for infra and design.}} \\[3pt]
Yalin Li,
Jie Fang,
Yifan Duan,
Zeyue Tian,
Guangzheng Li,
Haina Zhu,
Shuyi Wang, 
Jinwen Wang,
Mingyu Cui,
Tian Tan,
Auden,
Sen Liang

\vspace{0.4cm}

\noindent{\faHandshake\;\textcolor{black}{\textbf{Project Sponsors \& Advisors}}} \\[4pt]
Steve Yves, 
Shan Yang, 
Liefeng Bo,
Zilong Zheng,
Kai Yu,
Eng-Siong Chng, 
Xie  Chen\textsuperscript{*}\begingroup\def\thefootnote{*}\footnotetext[0]{Xie Chen is the corresponding author.}\endgroup

\vspace{0.4cm}

\noindent{\faHeart\;\textcolor{black}{\textbf{Acknowledgements}}} \\[4pt]
Yushen Chen,
Wenxi Chen,
Feiteng Li,
Qixi Zheng,
Yiwei Guo,
Guanrou Yang,
Yipeng Kang,
Shengpeng Ji,
Yuzhe Liang,
Peifan Chen,
Qixiang Xu,
Jiayi Liang,
Jubin Zhang,
Jiaxin Zhi,
Shanyi Zhu,
Yiru Fan,
Pan Luo

\clearpage
\bibliographystyle{plain}
\bibliography{references}

\clearpage
\appendix
\section{Detailed Evaluation Results}
\label{app:detailed-evaluation-results}

This appendix complements the benchmark summary in \Cref{tab:sota-summary} with detailed comparisons for \modelname and \modelname-Flash.
The evaluation is organized into speech generation, general instruction-guided editing, and signal-level enhancement and separation.
We retain each benchmark's native metrics and include broader baseline sets to compare linguistic accuracy, speaker preservation, instruction adherence, and perceptual quality.

\subsection{Generation Benchmarks}

We evaluate two complementary generation settings.
Seed-TTS-Eval~\cite{seedtts} measures zero-shot voice cloning across English, Chinese, and challenging Chinese text using recognition error and speaker similarity, whereas InstructTTSEval~\cite{instructttseval} measures control over acoustic parameters, descriptive styles, and role-playing instructions in Chinese and English.
For both benchmarks, we report the mean over three runs and use only benchmark-specific instruction templates and duration estimation, without prompt enhancement.

\FloatBarrier
\begin{table*}[!htbp]
\centering
\footnotesize
\setlength{\tabcolsep}{4pt}
\renewcommand{\arraystretch}{1.08}
\caption{Zero-shot TTS performance on Seed-TTS-Eval.}
\label{tab:seed-tts-eval}
\begin{adjustbox}{width=\textwidth}
\begin{tabular}{lcccccccc}
\toprule
\multirow{2}{*}{\textbf{Model}} &
\multicolumn{2}{c}{\textbf{test-en}} &
\multicolumn{2}{c}{\textbf{test-zh}} &
\multicolumn{2}{c}{\textbf{test-zh-hard}} &
\multicolumn{2}{c}{\textbf{Average}} \\
\cmidrule(lr){2-3}\cmidrule(lr){4-5}\cmidrule(lr){6-7}\cmidrule(lr){8-9}
& \textbf{WER(\%)}$\downarrow$ & \textbf{SIM}$\uparrow$ & \textbf{CER(\%)}$\downarrow$ & \textbf{SIM}$\uparrow$ & \textbf{CER(\%)}$\downarrow$ & \textbf{SIM}$\uparrow$ & \textbf{CER/WER(\%)}$\downarrow$ & \textbf{SIM}$\uparrow$ \\
\midrule
F5-TTS~\cite{f5tts} & 2.00 & 0.670 & 1.53 & 0.760 & 8.67 & 0.713 & 4.10 & 0.714 \\
FireRedTTS 2~\cite{xie2025fireredtts2} & 1.95 & 0.665 & 1.14 & 0.736 & 8.98 & 0.703 & 4.02 & 0.701 \\
IndexTTS 2~\cite{zhou2026indextts2} & 2.23 & 0.706 & 1.03 & 0.765 & 7.12 & 0.755 & 3.46 & 0.742 \\
MegaTTS 3~\cite{jiang2025megatts} & 2.79 & 0.771 & 1.52 & 0.790 & -- & -- & -- & -- \\
Qwen3-TTS~\cite{qwen3tts} & 1.23 & 0.717 & 1.22 & 0.770 & 6.76 & 0.748 & 3.07 & 0.745 \\
Seed-TTS~\cite{seedtts} & 2.25 & 0.762 & 1.12 & \underline{0.796} & 7.59 & 0.776 & 3.65 & 0.778 \\
DiTAR~\cite{jia2025ditar} & 1.69 & 0.735 & \underline{1.02} & 0.753 & -- & -- & -- & -- \\
VibeVoice~\cite{peng2025vibevoice} & 3.04 & 0.689 & 1.16 & 0.744 & -- & -- & -- & -- \\
VoxCPM 2~\cite{voxcpm2} & 1.84 & 0.753 & \textbf{0.97} & 0.795 & 8.13 & 0.753 & 3.65 & 0.767 \\
\midrule
\modelname-Flash & \underline{1.03} & \underline{0.772} & 1.10 & \textbf{0.814} & \underline{6.43} & \textbf{0.784} & \underline{2.85} & \underline{0.790} \\
\modelname & \textbf{1.02} & \textbf{0.788} & \underline{1.02} & \textbf{0.814} & \textbf{5.91} & \underline{0.782} & \textbf{2.65} & \textbf{0.795} \\
\bottomrule
\end{tabular}
\end{adjustbox}
\end{table*}

\begin{table*}[!htbp]
\centering
\footnotesize
\setlength{\tabcolsep}{4pt}
\renewcommand{\arraystretch}{1.08}
\caption{Instruction-following TTS performance on InstructTTSEval.}
\label{tab:instruct-tts-eval}
\begin{adjustbox}{max width=\textwidth}
\begin{tabular}{lcccccc}
\toprule
\multirow{2}{*}{\textbf{Model}} &
\multicolumn{3}{c}{\textbf{Chinese}} &
\multicolumn{3}{c}{\textbf{English}} \\
\cmidrule(lr){2-4}\cmidrule(lr){5-7}
& \textbf{APS}$\uparrow$ & \textbf{DSD}$\uparrow$ & \textbf{RP}$\uparrow$ & \textbf{APS}$\uparrow$ & \textbf{DSD}$\uparrow$ & \textbf{RP}$\uparrow$ \\
\midrule
Qwen3-TTS-VD~\cite{qwen3tts} & \textbf{85.20} & \underline{81.10} & 65.10 & \underline{82.90} & \textbf{82.40} & \underline{68.40} \\
Mimo-Audio~\cite{mimoaudio} & 75.70 & 74.30 & 61.50 & 80.60 & 77.60 & 59.50 \\
VoiceSculptor~\cite{hu2026voicesculptor} & 75.70 & 64.70 & 61.50 & -- & -- & -- \\
Hume AI~\cite{humeai2026} & -- & -- & -- & \textbf{83.00} & 75.30 & 54.30 \\
VoxInstruct~\cite{zhou2024voxinstruct} & 47.50 & 52.30 & 42.60 & 54.90 & 57.00 & 39.30 \\
Parler-TTS-Mini~\cite{lyth2024natural} & -- & -- & -- & 63.40 & 48.70 & 28.60 \\
Parler-TTS-Large~\cite{lyth2024natural} & -- & -- & -- & 60.00 & 45.90 & 31.20 \\
PromptTTS~\cite{guo2023prompttts} & -- & -- & -- & 64.30 & 47.20 & 31.40 \\
PromptStyle~\cite{liu2023promptstyle} & -- & -- & -- & 57.40 & 46.40 & 30.90 \\
MOSS-VoiceGenerator~\cite{mossvoicegenerator} & 78.00 & 80.00 & \textbf{74.00} & 68.20 & \underline{82.00} & \textbf{68.70} \\
\midrule
\modelname-Flash & 81.78 & 78.80 & 63.13 & 72.33 & \textbf{82.40} & 63.87 \\
\modelname & \underline{83.28} & \textbf{83.37} & \underline{68.23} & 75.57 & 81.60 & 65.93 \\
\bottomrule
\end{tabular}
\end{adjustbox}
\end{table*}

\clearpage
\subsection{General Speech Editing Benchmarks}

General speech editing evaluates whether a model follows free-form editing instructions while preserving content and speaker attributes outside the requested change.
MMAE-Speech~\cite{ma2026mmae} measures instruction following, content retention, and overall edit success with the Prompt Enhancer enabled.
Ming-Freeform-Audio-Edit~\cite{yan2025ming} further decomposes semantic editing into deletion, insertion, and substitution under bilingual Basic and Full settings, and evaluates acoustic control over speaking rate, pitch, and volume.
The detailed tables report operation-level results rather than only the averages presented in \Cref{tab:sota-summary}.

\FloatBarrier
\begin{table*}[!htbp]
\centering
\footnotesize
\setlength{\tabcolsep}{4pt}
\renewcommand{\arraystretch}{1.08}
\caption{Instruction-guided speech editing performance on MMAE-Speech with the Prompt Enhancer enabled. Models marked with $^*$ are evaluated only on MMAE samples with input durations of at most 10 seconds, following the original benchmark protocol.}
\label{tab:mmae-speech}
\begin{tabular}{lccc}
\toprule
\textbf{Model} & \textbf{IFR}$\uparrow$ & \textbf{CR}$\uparrow$ & \textbf{EMR}$\uparrow$ \\
\midrule
Step-Audio-EditX~\cite{yan2025step} & 43.52 & 77.27 & 4.69 \\
Ming-UniAudio~\cite{yan2025ming} & 34.13 & 76.01 & 7.04 \\
MMEdit$^*$~\cite{mmedit} & 30.52 & 35.40 & 0.99 \\
Audio-Omni$^*$~\cite{audio-omni} & 43.14 & 68.29 & 1.98 \\
SmartDJ$^*$ w/o planner~\cite{lan2025guiding} & 28.03 & 56.22 & 2.97 \\
SmartDJ$^*$ w/ planner~\cite{lan2025guiding} & 32.17 & 43.00 & 0.99 \\
\midrule
\modelname-Flash & \underline{46.62} & \underline{86.41} & \textbf{13.85} \\
\modelname & \textbf{48.23} & \textbf{88.11} & \underline{12.44} \\
\bottomrule
\end{tabular}
\end{table*}

\begin{table*}[!htbp]
\centering
\scriptsize
\setlength{\tabcolsep}{2.5pt}
\renewcommand{\arraystretch}{1.08}
\caption{Semantic speech editing performance on the Basic split of Ming-Freeform-Audio-Edit.}
\label{tab:ming-freeform-semantic-basic}
\begin{adjustbox}{max width=\textwidth}
\begin{tabular}{llcccccccccccc}
\toprule
\multirow{2}{*}{\textbf{Lang.}} &
\multirow{2}{*}{\textbf{Model}} &
\multicolumn{4}{c}{\textbf{Deletion}} &
\multicolumn{4}{c}{\textbf{Insertion}} &
\multicolumn{4}{c}{\textbf{Substitution}} \\
\cmidrule(lr){3-6}\cmidrule(lr){7-10}\cmidrule(lr){11-14}
& & \textbf{WER(\%)}$\downarrow$ & \textbf{ACC}$\uparrow$ & \textbf{SIM}$\uparrow$ & \textbf{no-edit WER(\%)}$\downarrow$
& \textbf{WER(\%)}$\downarrow$ & \textbf{ACC}$\uparrow$ & \textbf{SIM}$\uparrow$ & \textbf{no-edit WER(\%)}$\downarrow$
& \textbf{WER(\%)}$\downarrow$ & \textbf{ACC}$\uparrow$ & \textbf{SIM}$\uparrow$ & \textbf{no-edit WER(\%)}$\downarrow$ \\
\midrule
\multirow{3}{*}{ZH}
& Ming-UniAudio~\cite{yan2025ming} & 11.89 & \textbf{100.00} & \underline{0.78} & 11.49 & \underline{3.42} & 80.00 & 0.83 & \underline{3.52} & 4.52 & 78.62 & \underline{0.82} & 4.63 \\
& \modelname-Flash & \underline{9.85} & \textbf{100.00} & \textbf{0.80} & \underline{9.68} & 3.60 & \underline{82.94} & \textbf{0.93} & 3.96 & \textbf{1.15} & \textbf{93.08} & \textbf{0.91} & \textbf{1.38} \\
& \modelname & \textbf{9.61} & \textbf{100.00} & \textbf{0.80} & \textbf{9.46} & \textbf{3.03} & \textbf{85.29} & \underline{0.92} & \textbf{3.38} & \underline{1.21} & \underline{90.57} & \textbf{0.91} & \underline{1.46} \\
\midrule
\multirow{3}{*}{EN}
& Ming-UniAudio~\cite{yan2025ming} & 14.85 & \underline{82.22} & 0.76 & 24.26 & 6.63 & 71.43 & 0.79 & 17.70 & 8.99 & 59.78 & \underline{0.78} & 19.28 \\
& \modelname-Flash & \underline{7.38} & \textbf{100.00} & \textbf{0.84} & \underline{18.98} & \underline{4.26} & \underline{78.26} & \textbf{0.93} & \underline{15.61} & \underline{2.48} & \textbf{73.74} & \textbf{0.89} & \underline{15.06} \\
& \modelname & \textbf{5.91} & \textbf{100.00} & \underline{0.83} & \textbf{17.76} & \textbf{3.47} & \textbf{83.23} & \underline{0.92} & \textbf{15.01} & \textbf{2.18} & \underline{73.18} & \textbf{0.89} & \textbf{14.74} \\
\bottomrule
\end{tabular}
\end{adjustbox}
\end{table*}

\begin{table*}[!htbp]
\centering
\scriptsize
\setlength{\tabcolsep}{2.5pt}
\renewcommand{\arraystretch}{1.08}
\caption{Semantic speech editing performance on the Full split of Ming-Freeform-Audio-Edit.}
\label{tab:ming-freeform-semantic-full}
\begin{adjustbox}{max width=\textwidth}
\begin{tabular}{llcccccccccccc}
\toprule
\multirow{2}{*}{\textbf{Lang.}} &
\multirow{2}{*}{\textbf{Model}} &
\multicolumn{4}{c}{\textbf{Deletion}} &
\multicolumn{4}{c}{\textbf{Insertion}} &
\multicolumn{4}{c}{\textbf{Substitution}} \\
\cmidrule(lr){3-6}\cmidrule(lr){7-10}\cmidrule(lr){11-14}
& & \textbf{WER(\%)}$\downarrow$ & \textbf{ACC}$\uparrow$ & \textbf{SIM}$\uparrow$ & \textbf{no-edit WER(\%)}$\downarrow$
& \textbf{WER(\%)}$\downarrow$ & \textbf{ACC}$\uparrow$ & \textbf{SIM}$\uparrow$ & \textbf{no-edit WER(\%)}$\downarrow$
& \textbf{WER(\%)}$\downarrow$ & \textbf{ACC}$\uparrow$ & \textbf{SIM}$\uparrow$ & \textbf{no-edit WER(\%)}$\downarrow$ \\
\midrule
\multirow{3}{*}{ZH}
& Ming-UniAudio~\cite{yan2025ming} & 22.92 & 82.92 & 0.81 & 17.50 & 3.89 & 79.31 & 0.83 & 4.10 & 4.56 & 76.62 & 0.83 & 4.75 \\
& \modelname-Flash & \underline{5.17} & \underline{98.22} & \textbf{0.83} & \underline{4.50} & \underline{3.04} & \underline{85.86} & \textbf{0.93} & \underline{3.40} & \underline{1.82} & \textbf{89.54} & \textbf{0.92} & \underline{2.07} \\
& \modelname & \textbf{4.75} & \textbf{98.58} & \underline{0.82} & \textbf{4.20} & \textbf{2.77} & \textbf{86.90} & \underline{0.92} & \textbf{3.14} & \textbf{1.75} & \underline{88.92} & \underline{0.91} & \textbf{1.99} \\
\midrule
\multirow{3}{*}{EN}
& Ming-UniAudio~\cite{yan2025ming} & 27.60 & \underline{85.00} & 0.74 & 35.21 & 7.59 & 62.31 & 0.79 & 18.84 & 7.64 & 65.62 & \underline{0.77} & 18.39 \\
& \modelname-Flash & \underline{7.55} & \textbf{100.00} & \textbf{0.84} & \underline{18.90} & \underline{4.29} & \underline{75.88} & \textbf{0.93} & \underline{15.95} & \underline{2.68} & \underline{73.05} & \textbf{0.89} & \underline{15.44} \\
& \modelname & \textbf{6.17} & \textbf{100.00} & \underline{0.83} & \textbf{17.81} & \textbf{3.38} & \textbf{81.91} & \underline{0.92} & \textbf{15.09} & \textbf{2.34} & \textbf{73.83} & \textbf{0.89} & \textbf{15.06} \\
\bottomrule
\end{tabular}
\end{adjustbox}
\end{table*}

\begin{table*}[!htbp]
\centering
\footnotesize
\setlength{\tabcolsep}{5pt}
\renewcommand{\arraystretch}{1.08}
\caption{Acoustic speech editing performance on Ming-Freeform-Audio-Edit. RDE and RAE are reported for speed and volume alteration, respectively; bold and underlined values denote the best and second-best result for each language and task.}
\label{tab:ming-freeform-acoustic}
\begin{adjustbox}{max width=\textwidth}
\begin{tabular}{llcccccccc}
\toprule
\multirow{2}{*}{\textbf{Lang.}} &
\multirow{2}{*}{\textbf{Model}} &
\multicolumn{3}{c}{\textbf{Speed Alteration}} &
\multicolumn{2}{c}{\textbf{Pitch Alteration}} &
\multicolumn{3}{c}{\textbf{Volume Alteration}} \\
\cmidrule(lr){3-5}\cmidrule(lr){6-7}\cmidrule(lr){8-10}
& & \textbf{WER(\%)}$\downarrow$ & \textbf{SIM}$\uparrow$ & \textbf{RDE(\%)}$\downarrow$
& \textbf{WER(\%)}$\downarrow$ & \textbf{SIM}$\uparrow$
& \textbf{WER(\%)}$\downarrow$ & \textbf{SIM}$\uparrow$ & \textbf{RAE(\%)}$\downarrow$ \\
\midrule
\multirow{3}{*}{ZH}
& Ming-UniAudio~\cite{yan2025ming} & 5.88 & 0.66 & \textbf{6.36} & 7.45 & \underline{0.36} & \underline{1.71} & 0.86 & \textbf{14.90} \\
& \modelname-Flash & \textbf{2.60} & \textbf{0.82} & \underline{10.57} & \underline{2.67} & \textbf{0.58} & 1.93 & \textbf{0.97} & \underline{24.69} \\
& \modelname & \underline{2.72} & \underline{0.79} & \underline{10.57} & \textbf{1.82} & \textbf{0.58} & \textbf{1.52} & \underline{0.96} & 32.89 \\
\midrule
\multirow{3}{*}{EN}
& Ming-UniAudio~\cite{yan2025ming} & 17.53 & 0.57 & \textbf{5.92} & 13.37 & \underline{0.24} & \textbf{1.35} & 0.80 & \textbf{11.70} \\
& \modelname-Flash & \underline{5.66} & \textbf{0.76} & \underline{11.96} & \underline{4.94} & \textbf{0.53} & 2.76 & \textbf{0.96} & \underline{18.47} \\
& \modelname & \textbf{4.80} & \underline{0.75} & \underline{11.96} & \textbf{3.92} & \textbf{0.53} & \underline{1.73} & \underline{0.95} & 25.40 \\
\bottomrule
\end{tabular}
\end{adjustbox}
\end{table*}

\clearpage
\subsection{Speech Enhancement and Separation Benchmarks}

These benchmarks assess signal-level processing across denoising, separation, and super-resolution.
DNS Challenge 2020~\cite{dnschallenge2020} and CHiME-4~\cite{chime4} evaluate speech enhancement, Libri2Mix~\cite{libri2mix} evaluates two-speaker separation, and VCTK-SR~\cite{vctksr} evaluates recovery from bandwidth and channel degradations.
DNSMOS and UTMOS estimate perceptual quality; WER, dWER, and PER measure linguistic preservation; and SIM measures speaker-identity preservation.
Reporting these metrics together distinguishes perceptual improvement from content loss or speaker drift.

\FloatBarrier
\begin{table*}[!htbp]
\centering
\scriptsize
\setlength{\tabcolsep}{3pt}
\renewcommand{\arraystretch}{1.08}
\caption{Speech enhancement performance on DNS Challenge 2020.}
\label{tab:dnschallenge-complete}
\begin{adjustbox}{max width=\textwidth}
\begin{tabular}{lcccccc}
\toprule
\multirow{2}{*}{\textbf{Model}} & \multicolumn{3}{c}{\textbf{DNSMOS}} & \multirow{2}{*}{\textbf{UTMOS}$\uparrow$} & \multirow{2}{*}{\textbf{dWER(\%)}$\downarrow$} & \multirow{2}{*}{\textbf{SIM}$\uparrow$} \\
\cmidrule(lr){2-4}
& \textbf{OVRL}$\uparrow$ & \textbf{SIG}$\uparrow$ & \textbf{BAK}$\uparrow$ & & & \\
\midrule
Ming-UniAudio~\cite{yan2025ming} & 3.26 & 3.57 & 3.99 & 3.73 & 6.91 & 0.97 \\
Resemble-Enhance~\cite{resemble-enhance} & 3.37 & \underline{3.62} & 4.12 & 3.49 & 6.89 & 0.97 \\
DaSheng~\cite{dasheng} & 3.37 & 3.58 & \underline{4.17} & 3.46 & 3.48 & \underline{0.98} \\
SAM-Audio-Large~\cite{samaudio} & 3.29 & 3.56 & 4.06 & 3.78 & 9.67 & 0.96 \\
AnyEnhance~\cite{zhang2025anyenhance} & \textbf{3.41} & \textbf{3.63} & \textbf{4.19} & 3.76 & 7.71 & \underline{0.98} \\
RE-USE~\cite{fu2026rethinking} & \underline{3.38} & 3.60 & \textbf{4.19} & 3.69 & 3.31 & \underline{0.98} \\
\midrule
\modelname-Flash & \underline{3.38} & \underline{3.62} & 4.13 & \textbf{4.05} & \underline{2.93} & \textbf{0.99} \\
\modelname & 3.35 & 3.59 & 4.13 & \underline{3.86} & \textbf{2.66} & \textbf{0.99} \\
\bottomrule
\end{tabular}
\end{adjustbox}
\end{table*}

\begin{table*}[!htbp]
\centering
\scriptsize
\setlength{\tabcolsep}{4pt}
\renewcommand{\arraystretch}{1.08}
\caption{Speech enhancement performance on CHiME-4.}
\label{tab:chime4-complete}
\begin{adjustbox}{max width=\textwidth}
\begin{tabular}{lccccc}
\toprule
\multirow{2}{*}{\textbf{Model}} & \multicolumn{3}{c}{\textbf{DNSMOS}} & \multirow{2}{*}{\textbf{UTMOS}$\uparrow$} & \multirow{2}{*}{\textbf{WER(\%)}$\downarrow$} \\
\cmidrule(lr){2-4}
& \textbf{OVRL}$\uparrow$ & \textbf{SIG}$\uparrow$ & \textbf{BAK}$\uparrow$ & & \\
\midrule
Ming-UniAudio~\cite{yan2025ming} & 3.19 & 3.47 & 4.04 & 3.56 & 21.66 \\
Resemble-Enhance~\cite{resemble-enhance} & \textbf{3.35} & \textbf{3.60} & 4.12 & 3.36 & 31.90 \\
DaSheng~\cite{dasheng} & 3.27 & 3.49 & \underline{4.17} & 3.34 & 10.89 \\
SAM-Audio-Large~\cite{samaudio} & 3.14 & 3.49 & 3.90 & 3.38 & 17.30 \\
AnyEnhance~\cite{zhang2025anyenhance} & 3.19 & 3.41 & 4.13 & 3.24 & 27.41 \\
RE-USE~\cite{fu2026rethinking} & \underline{3.33} & 3.55 & \textbf{4.18} & 3.44 & 10.71 \\
\midrule
\modelname-Flash & \textbf{3.35} & \underline{3.59} & 4.14 & \textbf{3.91} & \textbf{7.84} \\
\modelname & 3.28 & 3.51 & 4.14 & \underline{3.72} & \underline{7.98} \\
\bottomrule
\end{tabular}
\end{adjustbox}
\end{table*}

\begin{table*}[!htbp]
\centering
\scriptsize
\setlength{\tabcolsep}{4pt}
\renewcommand{\arraystretch}{1.08}
\caption{Two-speaker speech separation performance on Libri2Mix.}
\label{tab:libri2mix-complete}
\begin{adjustbox}{max width=\textwidth}
\begin{tabular}{lccccccc}
\toprule
\multirow{2}{*}{\textbf{Model}} & \multicolumn{3}{c}{\textbf{DNSMOS}} & \multirow{2}{*}{\textbf{UTMOS}$\uparrow$} & \multirow{2}{*}{\textbf{WER(\%)}$\downarrow$} & \multirow{2}{*}{\textbf{PER(\%)}$\downarrow$} & \multirow{2}{*}{\textbf{SIM}$\uparrow$} \\
\cmidrule(lr){2-4}
& \textbf{OVRL}$\uparrow$ & \textbf{SIG}$\uparrow$ & \textbf{BAK}$\uparrow$ & & & & \\
\midrule
MossFormer2-SS~\cite{mossformer2} & 3.24 & 3.48 & \underline{4.11} & 3.66 & \underline{9.34} & \underline{7.23} & \textbf{0.96} \\
SAM-Audio-Large~\cite{samaudio} & 2.87 & 3.27 & 3.55 & 2.84 & 57.22 & 47.00 & 0.86 \\
Sidon (Dialogue)~\cite{dialoguesidon} & 3.08 & 3.48 & 3.77 & 2.61 & 51.79 & 39.57 & \underline{0.87} \\
\midrule
\modelname-Flash & \textbf{3.32} & \textbf{3.57} & \textbf{4.15} & \textbf{4.03} & 10.07 & 7.62 & \textbf{0.96} \\
\modelname & \underline{3.28} & \underline{3.52} & \textbf{4.15} & \underline{3.87} & \textbf{9.12} & \textbf{6.63} & \textbf{0.96} \\
\bottomrule
\end{tabular}
\end{adjustbox}
\end{table*}

\begin{table*}[!htbp]
\centering
\scriptsize
\setlength{\tabcolsep}{4pt}
\renewcommand{\arraystretch}{1.08}
\caption{Speech super-resolution performance on VCTK-SR.}
\label{tab:vctk-sr-complete}
\begin{adjustbox}{max width=\textwidth}
\begin{tabular}{lcccccccc}
\toprule
\multirow{2}{*}{\textbf{Model}} & \multicolumn{3}{c}{\textbf{DNSMOS}} & \multirow{2}{*}{\textbf{UTMOS}$\uparrow$} & \multirow{2}{*}{\textbf{WER(\%)}$\downarrow$} & \multirow{2}{*}{\textbf{PER(\%)}$\downarrow$} & \multirow{2}{*}{\textbf{SIM}$\uparrow$} \\
\cmidrule(lr){2-4}
& \textbf{OVRL}$\uparrow$ & \textbf{SIG}$\uparrow$ & \textbf{BAK}$\uparrow$ & & & & \\
\midrule
AudioSR~\cite{audiosr} & 3.08 & 3.39 & 4.00 & 3.13 & 4.72 & 20.51 & 0.92 \\
Resemble-Enhance~\cite{resemble-enhance} & 3.18 & 3.45 & \underline{4.09} & 3.59 & 15.94 & 19.61 & 0.95 \\
\midrule
\modelname-Flash & \textbf{3.25} & \textbf{3.51} & \textbf{4.12} & \textbf{4.05} & \textbf{2.92} & \textbf{4.19} & \underline{0.96} \\
\modelname & \underline{3.21} & \underline{3.47} & 4.08 & \underline{3.93} & \underline{3.06} & \underline{4.72} & \textbf{0.97} \\
\bottomrule
\end{tabular}
\end{adjustbox}
\end{table*}

\FloatBarrier
\end{document}